\documentclass[aps,prb,floatfix,twocolumn,superscriptaddress]{revtex4-2}
\usepackage[utf8]{inputenc}
\usepackage{mathrsfs}
\usepackage{graphicx}
\usepackage{epstopdf}
\usepackage[english]{babel}
\usepackage{amsmath}
\usepackage{amssymb}
\usepackage{xcolor}
\usepackage{relsize}
\usepackage{CJK}
\usepackage{textcomp}
\usepackage{dsfont}
\usepackage{bm}
\usepackage{float}
\renewcommand{\parallel}{\mathrel{/\mskip-4mu/}}

\newcommand{\me}{\mathrm{e}}      
\newcommand{\mi}{\mathrm{i}}     

\newcommand{\dif}{\mathrm{d}}

\allowdisplaybreaks

\begin{document}

\title{Geometry near rank-changing points on the mixed-state manifold: Bures metric, conical singularities, and Lindblad dynamics}
\author{Yu-Huan Huang}
\affiliation{School of Physics, Southeast University, Jiulonghu Campus, Nanjing 211189, China}
\author{Xu-Yang Hou}
\affiliation{School of Physics, Southeast University, Jiulonghu Campus, Nanjing 211189, China}

\author{Hao Guo}
\email{guohao.ph@seu.edu.cn}
\affiliation{School of Physics, Southeast University, Jiulonghu Campus, Nanjing 211189, China}
\affiliation{Hefei National Laboratory, University of Science and Technology of China, Hefei 230088, China}

\author{Chih-Chun Chien}
\email{cchien5@ucmerced.edu}
\affiliation{Department of physics, University of California, Merced, CA 95343, USA}

\begin{abstract}
We elucidate the Bures metric in quantum state space near a rank-changing point of the density matrix and show contrasting behavior for two-level ($N=2$) systems versus higher-level systems. Due to the smooth pure-state boundary for $N=2$, we prove the apparent metric divergences to be merely coordinate artifacts and present three Lindblad processes exhibiting qualitatively different evolution near rank-changing points, showing geodesic approach, power-law scaling, and pure-state escape law.
For higher-dimensional ($N\ge 3$) systems, the geometry near a rank-changing point differs fundamentally. Under suitable restrictions of the density matrix and its approach towards a pure state,
the Bures metric reduces to a conical metric with the pure state at the cone tip. 
Such a conic geometry leads to genuine curvature singularities: A two-dimensional cone exhibits a Dirac delta-function curvature near the tip while a higher-dimensional cone shows a power-law divergence of the curvature towards the cone tip. A construction of Lindblad evolution for $N=3$ systems with conic singularities is presented, along with possible implications for future experimental and theoretical research. 
\end{abstract}

\maketitle
		
\section{Introduction}
The geometry of quantum states~\cite{Bengtsson_book,doi:10.1142/S0129055X20300010,10.1093/nsr/nwae334,Yu25,Verma2026} helps elucidate important conceptual underpinnings of quantum information and condensed matter. The Bures metric~\cite{Bures1969} serves as the natural Riemannian structure on mixed states and provides a measure of statistical distinguishability. Moreover, it connects to quantum Fisher information and has potential applications in metrology~\cite{PhysRevLett.72.3439,Liu_2020}. The Bures metric is part of a fundamental object, called the quantum geometric tensor (QGT), which characterizes the local geometry of quantum states under parameter variations~\cite{QGTCMP80, QGT10, Berry84, Simon83, KOLODRUBETZ20171}. For pure states, the real and imaginary parts are respectively associated with the Fubini-Study metric and Berry curvature, linking geometry to topology~\cite{EGUCHI1980213, Berry84, Simon83}. The QGT has been extensively studied in quantum information, condensed matter, and AMO physics~\cite{Bengtsson_book, Bohm03, KOLODRUBETZ20171, QGTCMP80, RevModPhys.82.1959}, and experiments have successfully probed many aspects of the QGT using various platforms~\cite{10.1093/nsr/nwz193, PhysRevLett.122.210401, Yi23, PhysRevB.108.094508, PhysRevB.87.245103, PhysRevLett.124.197002, PhysRevX.10.041041}.

However, most of the studies focus on pure states, whose manifold is smooth and well-defined. For mixed states described by density matrices, the QGT acquires a richer structure. Notably, the U$(N)$ gauge-invariant real part of the QGT for full-rank density matrices coincides with the Bures metric~\cite{PhysRevB.110.035144}. Yet the mixed-state manifold is fundamentally different because it is a stratified space. When the number of non-vanishing eigenvalue of the density matrix changes, the rank of the density matrix varies accordingly. Examples include, but are not limited to, evolution from a pure state into a mixed state or vice versa.
In this work, we go beyond the QGT to investigate the behavior of the Bures metric and its curvature, which can become singular at those rank-changing points if proper conditions are satisfied. Near a pure state when all but one eigenvalues of the density matrix vanish, the geometry can develop genuine singularities when the Hilbert space dimension is above two.

The geometric structure of mixed quantum states is considerably more intricate than that of pure states. While pure states form a smooth Riemannian manifold equipped with the Fubini-Study metric, the space of mixed density matrices constitutes a convex set rather than a manifold in the conventional sense~\cite{Bengtsson_book}. Consequently, no globally defined metric can smoothly parametrize the entire mixed-state space towards pure states without encountering singularities. Surprisingly. two-level systems (with Hilbert-space dimension $N=2$) stand as a remarkable exception, where the Bloch-ball representation of mixed states endows the state space with a genuine manifold structure. Further, the Bloch sphere of pure states forms the boundary of the Bloch ball. For $N\ge 3$ systems, however, such simplicity is lost, and any metric structure imposed on the state space inevitably develops singular behavior at rank-changing points.

To concretely demonstrate these geometric singularities and prove them physically tangible, we engineer specific dynamical evolution via the Lindblad master equation~\cite{Lindblad1976,10.1063/1.5115323}. The exemplary systems are driven through rank-changing transitions to reveal singular behaviors of local geometry. For $N=2$ systems, 
we demonstrate that apparent divergences in the Bures metric components at the boundary arise only from the coordinate
choice. Through appropriate coordinate transformations, the pure-state boundary reveals itself as a smooth, non-singular surface where the induced metric reduces to the Fubini-Study metric on the pure-state manifold. The curvature from the metric remains a finite constant, including the boundary of pure states, confirming that no genuine curvature singularity exists for $N=2$. We analyze three exemplary Lindblad evolution for $N=2$ systems to illustrate their reactions to a change of the rank of the density matrix.




However, the situation changes qualitatively for systems with higher-dimensional Hilbert space ($N\ge 3$) at rank-changing points. When the eigenvalues of the density matrix follow certain constraints, the Bures metric near a pure state develops genuine singularities. Further, under appropriate restrictions that isolate the angular degrees of freedom of the subspace orthogonal to the pure state of interest, the Bures metric reduces locally to a conical metric. Depending on the number of angular parameters in the metric cone, Dirac delta-function or power-law divergence in the scalar curvature associated with the Bures metric emerges at the cone tip.
We also construct a $N=3$ Lindblad process from pure to mixed states showing similar pure-state escape behavior, where the system follows radial motion near the cone tip, as well as a recipe for realizing the conditions for conic singularities through a Lindblad process. Our analyses of the Buries metric and its curvature thus reveal geometric singularities as systems approach rank-changing points of their density matrices.

The rest of the paper is organized as follows. Section \ref{Sec.3} introduces the geometric framework of the Bures metric and establishes the relevant notation, followed by an analysis of the $N=2$ case. Three concrete examples of Lindblad evolution are presented to visualize the impacts of rank-changing points. Section \ref{Sec.4} turns to $N\ge 3$ systems and presents their stratified structure. The conical metric reduction near a pure state is derived, and the curvature singularities are analyzed for different protocols with concrete examples from $N=3$ systems. Section~\ref{Sec:Implications} discusses possible implications for experimental and theoretical studies. Section \ref{Sec.5} concludes our work. The Appendix summarizes some details and proofs of the Buries metric and its curvature.


\section{Rank Changing for $N=2$ systems}\label{Sec.3}

\subsection{Overview of Bures Metric}\label{subsec:Bures}
We recall that for full-rank mixed states, the real part of the gauge-invariant quantum geometric tensor (QGT) reproduces the Bures metric~\cite{PhysRevB.110.035144} when restricting on the base manifold. The relation highlights the deep connection between the Bures geometry and the gauge-invariant structures underlying mixed states. The Bures distance between neighboring density matrices $\rho$ and $\rho+\dif\rho$ can be written as $\dif s_{\mathrm B}^2(\rho,\rho+\dif\rho)
	=
	g_{\mu\nu}\,\dif R^\mu \dif R^\nu$,
where the Bures metric tensor is
\begin{align}\label{Bmetric1}
	g^\mathrm{B}_{\mu\nu}
	=
	\frac12
	\sum_{i,j}
	\frac{\langle i|\partial_\mu\rho|j\rangle\langle j|\partial_\nu\rho|i\rangle}{\lambda_i+\lambda_j}.
\end{align}
Here $\{\lambda_i,|i\rangle\}$ denote the eigenvalues and eigenstates of $\rho$.

To investigate how the Bures metric behaves as the density matrix evolves on the mixed-state manifold, especially at points where the rank of the density matrix changes, we begin with the simplest nontrivial setting of two-level systems. For a two-level system, the Bloch vector parameterization maps the state space onto a three-dimensional ball, the Bloch ball, with the boundary formed by a two-dimensional sphere, the Bloch sphere, of pure states. Importantly, no conical singularities arise at the boundary for a two-level system in contrast to higher-dimensional systems, making the two-level system an ideal testing ground for studying metric behavior in the absence of intrinsic curvature singularities.

For two-level systems, $\rho$ can be written in Bloch form $\rho=\frac12\left(1+\boldsymbol{\sigma}\cdot\boldsymbol{x}\right)$, where $\boldsymbol{x}$ is the Bloch vector.
Let $r=|\boldsymbol{x}|$, $\boldsymbol{n}=\frac{\boldsymbol{x}}{r}$
and use $\boldsymbol{n}\cdot\dif\boldsymbol{n}=0$, then the Bures line element reduces to
\begin{align}\label{A13}
	\dif s_{\mathrm B}^2
	=
	\frac14\left(
	\frac{\dif r^2}{1-r^2}
	+
	r^2\,\dif\boldsymbol{n}\cdot\dif\boldsymbol{n}
	\right),
\end{align}
Taking further $\boldsymbol{n} = (\sin\theta\cos\phi,\sin\theta\sin\phi,\cos\theta)^T$, where $\theta$ and $\phi$ are the polar and azimuthal angles of the Bloch vector, respectively, Eq.~(\ref{A13}) leads to
\begin{align}\label{3}
	g_{rr}=\frac{1}{4}\frac{1}{1-r^2},\quad
	g_{\theta\theta}=\frac{1}{4}r^2,\quad
	g_{\phi\phi}=\frac{1}{4}r^2\sin^2\theta.
\end{align}
The Bloch representation offers a convenient visualization of the state, but the mapping $\rho\mapsto\boldsymbol{x}$ is not distance-preserving: an Euclidean straight line in the Bloch ball is not a Bures geodesic, and paths that appear simple in the Bloch picture may have nontrivial geometric properties under the Bures metric, as we will show below.

When $r=1$, the system becomes pure states, and Eq.~\eqref{A13} appears to diverge radially. To reveal the geometric nature of the pure states, we introduce $r=\cos \tilde{u}$ ($0\le \tilde{u}\le\pi/2$), which transforms Eq.~\eqref{A13} to
\begin{align}\label{A15}
\dif s_{\mathrm B}^2 = \frac14\bigl[\dif \tilde{u}^2+\cos^2\tilde{u}(\dif\theta^2+\sin^2\theta\dif\phi^2)\bigr].
\end{align}
At $\tilde{u}=0$ ($r=1$), this metric remains regular; restricting to the boundary $\dif \tilde{u}=0$ yields the induced metric
\begin{align}\label{dFS}
\dif s_{\mathrm{ind}}^2 = \frac14(\dif\theta^2+\sin^2\theta\dif\phi^2),
\end{align}
which is precisely the Fubini-Study metric of pure states~\cite{Bengtsson_book}. Hence the divergence at $r=1$ is merely a coordinate artifact, not an intrinsic curvature singularity, as a detailed calculation of the curvature from the Bures metric for $N=2$ systems in Appendix~\ref{app5} indeed shows no singularity.

We will construct three distinct Lindblad evolutions that drive the system through rank-changing transitions: (1) Asymptotic purification, where the system state approaches a pure state only in the infinite-time limit; (2) finite-time purification, where the rank of the density matrix drops from $2$ to $1$ at a specific finite time; (3) a decoherence process, in which a pure state evolves into a mixed state.
We will first present the time evolution of the Bures metrics for the three processes, then analyze their underlying geometric structure and provide a unified classification.

\subsection{Asymptotic Purification}

\begin{figure}[t]
\centering
\includegraphics[width=3.4in]{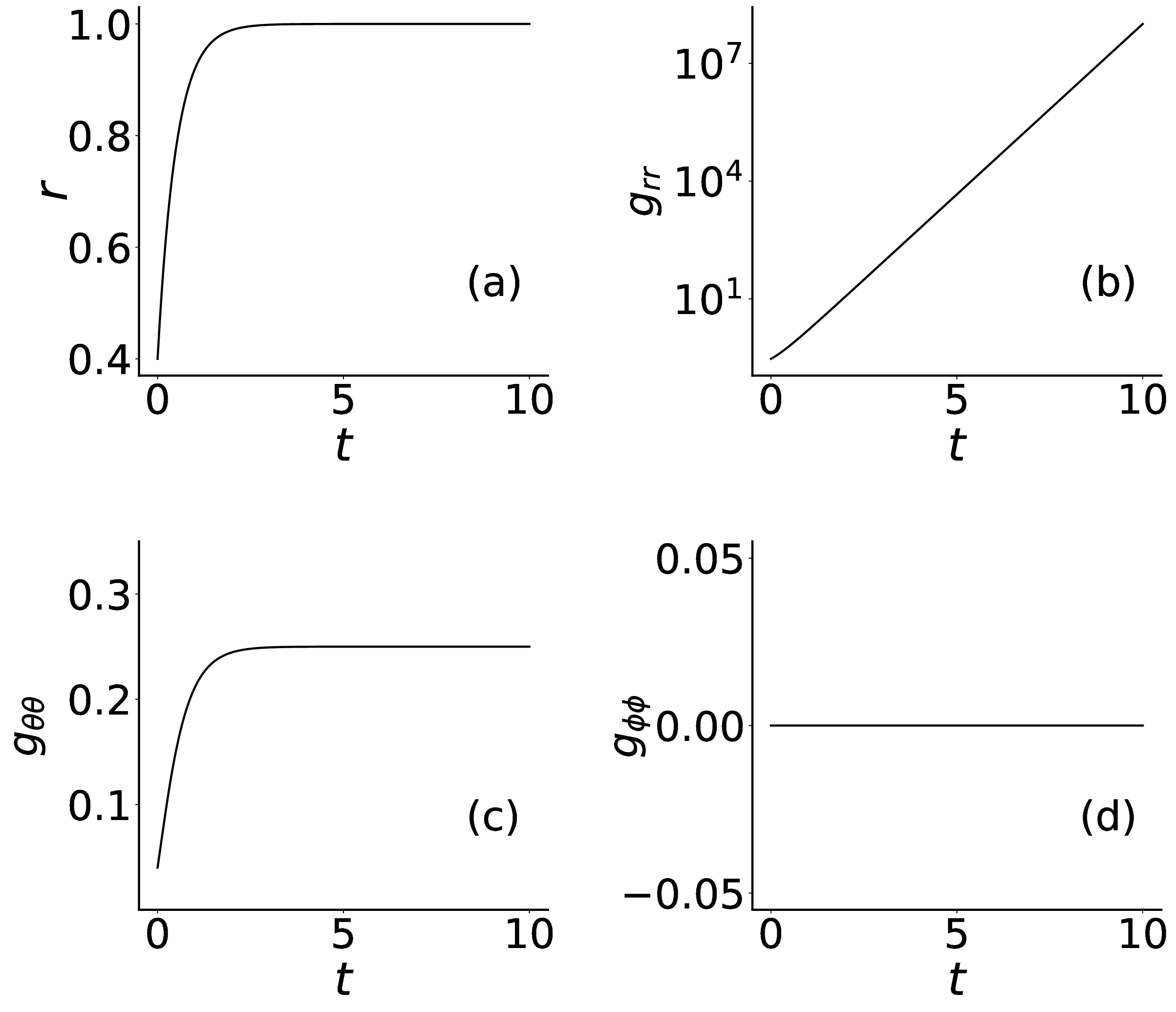}
\caption{(a) The Bloch vector norm $r$ and the Bures metric components (b) $g_{rr}, (C) g_{\theta\theta}, and (d) g_{\phi\phi}$ during the asymptotic purification evolution for $p=0.7$ and $\gamma=2.0$.}
\label{Fig1}
\end{figure}

We first consider the asymptotic purification of a two-level system, where an initial mixed state approaches a pure state only in the infinite-time limit. The basis states are denoted by $|0\rangle, |1\rangle$. We implement a time-independent Lindblad operator $L=\sqrt{\gamma}\,\sigma_-$ (with $\sigma_-=|0\rangle\langle 1|$) and set the Hamiltonian $H=0$ for simplicity. The initial state is chosen as a diagonal mixed state
\begin{align}
\rho(0)=p|0\rangle\langle0|+(1-p)|1\rangle\langle1|,\qquad 0<p<1.
\end{align}
We focus on the case $p>0.5$ so that the initial population in the ground state $|0\rangle$ is larger than that in $|1\rangle$; this ordering is preserved by the dynamics because $\rho_{00}(t)$ increases and $\rho_{11}(t)$ decreases monotonically. The Lindblad master equation~\cite{Lindblad1976}
\begin{align}\label{Lind}\displaystyle\frac{\dif\rho}{\dif t}=L\rho L^\dagger-\frac12\{L^\dagger L,\rho\}\end{align} then gives $\dot{\rho}_{00}=\gamma\rho_{11}$ and $\dot{\rho}_{11}=-\gamma\rho_{11}$,
where $\rho_{00}(t)+\rho_{11}(t)=1$. We assume the time has been normalized so $\gamma$ is dimensionless. Solving these equations yields
\begin{align}
\rho_{00}(t)=1-(1-p)\me^{-\gamma t},\quad
\rho_{11}(t)=(1-p)\me^{-\gamma t}.
\end{align}
As $t\to\infty$, $\rho(t)\to|0\rangle\langle0|$, so the system asymptotically approaches a pure state. Since $\rho_{00}(t)>\rho_{11}(t)$ for all $t$ when $p>0.5$, the smallest eigenvalue is $\lambda_{\min}(t)=\rho_{11}(t)=(1-p)\me^{-\gamma t}$, which decays exponentially. 

To analyze the Bures metric, we introduce the $t$-dependent Bloch vector $\boldsymbol{x}(t)=(x(t),y(t),z(t))^T$. Since the evolution preserves the diagonal form, we have
\begin{align}
x(t)=y(t)=0,\quad z(t)=1-2(1-p)\me^{-\gamma t},
\end{align}
so the radial coordinate $r(t)=|z(t)|=1-2\lambda_{\min}(t)$ with the Bloch vector fixed at $\theta=0$, i.e., along the $z$-axis. 
Hence changes in $\lambda_{\min}$ directly correspond to radial displacements.
Although the evolution appears as a simple radial line in the Bloch ball, the Bures geometry perceives it as a nontrivial curve. From Eq.~\eqref{3}, we obtain
\begin{align}
	g_{rr}(t)=\frac{1}{4[1-r(t)^2]},\quad g_{\theta\theta}(t)=\frac{r(t)^2}{4},\quad g_{\phi\phi}(t)=0.
\end{align}
As $t\to\infty$, $r\to1$, $\rho(t)\to|0\rangle\langle0|$, and the rank of the density matrix drops from $2$ to $1$. The angular component $g_{\theta\theta}$ tends to $1/4$, while the radial component $g_{rr}$ diverges exponentially because the denominator $1-r^2\sim 4\lambda_{\min}(t)\sim 4(1-p)\me^{-\gamma t}$ vanishes. This divergence originates from the apparent coordinate singularity at the pure-state boundary. These behaviors are illustrated in Figure \ref{Fig1}. However, no singularity occurs in the curvature associated with the Bures metric, as proven in Appendix~\ref{app5}.

In this amplitude damping process, 
the density matrix approaches the pure state $|0\rangle$ along the $z$-axis. The rank remains $2$ for all finite times and drops to $1$ only in the $t\to\infty$ limit. Consequently, the evolution stays entirely within the full-rank density matrices of the Bloch ball, approaching the boundary (the Bloch sphere) arbitrarily closely without ever touching it. Although the path appears trivial in the Bloch ball (simply moving along the $z$-axis), its geometric nature on the Bures manifold (the space of density matrices equipped with the Bures metric) is in fact a geodesic, i.e., a curve of shortest length connecting the initial state to the boundary.
To this end, we let $\eta=(1-p)\me^{-\gamma t}\in(0,1-p]$. Since the eigenstates of $\rho$ remain unchanged in this process, the Bures metric reduces to
\begin{align}
\dif s^2 = \frac{\dif \eta^2}{4x(1-\eta)}.
\end{align}
Introducing  $u=\arcsin\sqrt{\eta}$, then $\eta=\sin^2 u$, $\dif \eta=2\sin u\cos u\,\dif u$, and $\eta(1-\eta)=\sin^2 u\cos^2 u$. Using those expressions leads to $\dif s^2 = \dif u^2$, i.e., a flat metric. 
Meanwhile, geodesics satisfy $u(s)=s+s_0$ in these coordinates. 
Consequently, the trajectory is a geodesic on the Bures manifold.

This above special property of the amplitude damping process arises because the exponential decay $\eta\propto \me^{-\gamma t}$ coincides with the geodesic parameterization $\eta(s)=\sin^2(s+s_0)$ after flattening the coordinates. We caution that generic Lindblad evolutions typically do not follow this special condition. 
We emphasize the system approaches the rank-changing point radially, i.e., along the direction of the smallest eigenvalue $\lambda_{\min}$. This behavior, however, is more general as we will explain later.

\begin{figure}[t]
\centering
\includegraphics[width=3.4in]{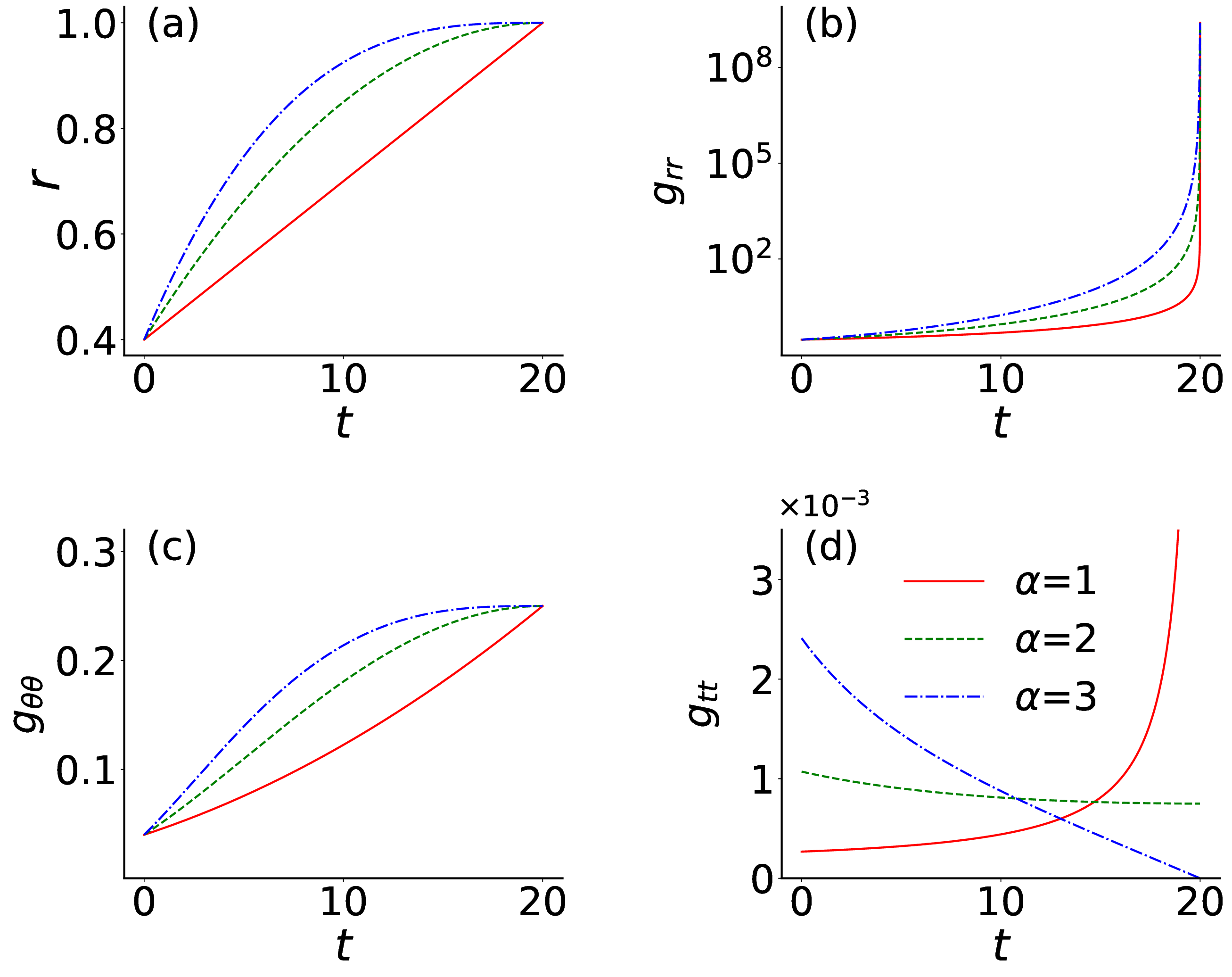}
\caption{Behaviors of (a) the Bloch vector norm and (b)-(d) the Bures metric components during the finite-time purification evolution for $p=0.7$, $\gamma=2.0$ and $\alpha=1.0$, 2.0, 3.0. Since $g_{\phi\phi}=0$ in this case, we show $g_{tt}$ in panel (d).}
\label{Fig2}
\end{figure}

\subsection{Finite-Time Purification}
We now turn to a second scenario where the system reaches a pure state in finite time, thereby undergoing a genuine rank change at a well-defined moment. To achieve this, a time-dependent decay rate is introduced so that the coefficient of the Lindblad operator diverges at the time when the system becomes a pure state. Specifically, we set
\begin{align}
\gamma(t)=\frac{\alpha}{T-t},\qquad 0\le t<T,\quad \alpha>0,
\end{align}
and take the Lindblad operator $L(t)=\sqrt{\gamma(t)}\,\sigma_-$. Using the same initial condition $\rho(0)=\operatorname{diag}(p,1-p)$  (with $p>0.5$ as before), the evolution equation (\ref{Lind}) yields
\begin{align}
\rho_{11}(t)=&(1-p)\left(\frac{T-t}{T}\right)^\alpha
\end{align}
and $\rho_{00}=1-\rho_{11}$. At $t=T$, $\rho(T)=|0\rangle\langle0|$, so the system is purified at finite time. The smallest eigenvalue for $t$ near $T$ is $\lambda_{\min}(t)=\rho_{11}(t)\sim (T-t)^\alpha$, which vanishes as a power law.

The Bloch vector also remains along the $z$-axis, with
\begin{align}
z(t)=1-2(1-p)\left(\frac{T-t}{T}\right)^\alpha,
\end{align}
so the radial coordinate is $r(t)=|z(t)|=1-2\lambda_{\min}(t)$. The Bures metric components are then
\begin{align}
g_{rr}(t)=&\frac{1}{4[1-r(t)^2]}\sim\frac{1}{16(1-p)}\left(\frac{T}{T-t}\right)^\alpha,\notag\\
g_{\theta\theta}(t)=&\frac{r(t)^2}{4},\qquad g_{\phi\phi}(t)=0.
\end{align}
As $t\to T$, the radial component $g_{rr}$ diverges following a power-law behavior, while the angular components remain finite. 
Figure \ref{Fig2} illustrates the dynamic characteristics of the Bloch vector norm and Bures metric components during this evolution for $p=0.7$ and $\alpha=1.0$, 2.0, 3.0. 
To understand the geometric meaning of the metric divergence, we note that because the Bloch vector remains fixed along the $z$-axis ($\theta=0$), the line element simplifies to $\dif s^2 = g_{rr}\,\dif r^2$, with $g_{rr}=1/[4(1-r^2)]$ from Eq.~\eqref{A13}. The induced metric component along the evolution direction is then $g_{tt}=g_{rr}\dot r^2$ (the angular terms do not contribute since $\dot\theta=0=\theta$). Using $\dot r\sim (T-t)^{\alpha-1}$ and $g_{rr}\sim (T-t)^{-\alpha}$, we obtain the universal scaling
\begin{align}
	g_{tt}\sim (T-t)^{\alpha-2}.
\end{align}
Thus $\alpha=2$ marks a critical point:
\begin{align}
	\alpha<2:&\quad g_{tt}\to\infty,\notag\\
	\alpha=2:&\quad g_{tt}\to\text{constant},\notag\\
	\alpha>2:&\quad g_{tt}\to 0.
\end{align}
As shown in Fig.~\ref{Fig2}(d), this behavior indicates a qualitative change in the stretching rate of the curve in Bures geometry at $\alpha=2$, which motivates the choices in Fig.~\ref{Fig2} to cover the three distinct regimes. 

Nevertheless, it is important to emphasize that this divergence remains a coordinate artifact arising from the $r$ coordinate on the Bloch ball. The transformation $r=\cos \tilde{u}$ regularizes the metric to the smooth form shown in Eq.~\eqref{A15}.
Moreover, although the evolution is also radial in Bloch coordinates similar to the asymptotic purification, it is generally not a geodesic of the Bures metric here. According to Eq.~(\ref{A13}), the Bures metric in Bloch coordinates for a two-level system is equivalent to
$\dif s^2 = \frac{\dif r^2}{4(1-r^2)} + \frac{r^2}{4}\dif\Omega^2$.
For a purely radial motion ($\dif\Omega=0$), its geodesic equation is given by $\ddot r(1-r^2) + r\dot r^2 = 0$, and the solution is
$r(t) = \sin(\omega t + \delta)$
with $\omega$ and $\delta$ being constants determined by the initial condition. In contrast, the finite-time purification process results in a power-law function $r(t)=1-2(1-p)[(T-t)/T]^\alpha$. 
Hence the evolution is not along a Bures geodesic but a specific path generated by the time-dependent Lindblad evolution, changing the rank at $t=T$. This is in stark contrast to the asymptotic purification case, where the path is a geodesic and the rank changes only in the limit $t\to\infty$.

\begin{figure}[t]
\centering
\includegraphics[width=\columnwidth]{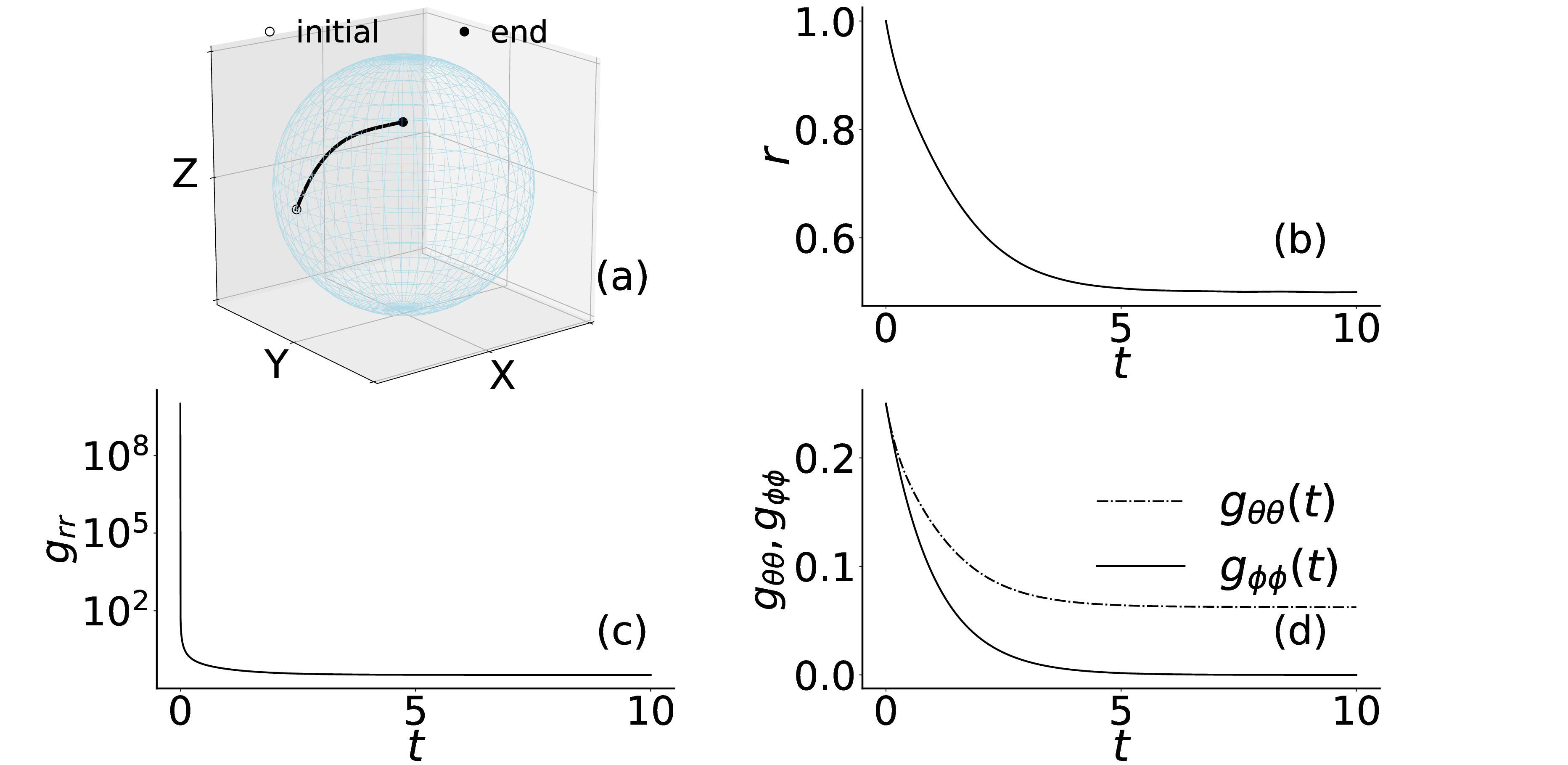}
\caption{(a) Evolution path from pure to mixed state in the Bloch ball. (b)-(d) Behavior of the Bloch vector norm and the Bures metric components during the evolution from pure to mixed state for $\gamma_1=1.0$ and $\gamma_2=0.5$.}
\label{Fig3}
\end{figure}

\subsection{Evolution from Pure to Mixed State}
We now examine the third process, in which a pure state decoheres into a mixed state. This illustrates how a system leaves the pure-state boundary of two-level systems. The initial state is chosen as the pure state $|+\rangle=\frac1{\sqrt2}(|0\rangle+|1\rangle)$, with  the density matrix
\begin{align}
\rho(0)=|+\rangle\langle+|=\frac12\begin{pmatrix}1&1\\1&1\end{pmatrix},
\end{align}
corresponding to the initial Bloch vector $\boldsymbol{x}(0)=(1,0,0)^T$.
Taking $H=0$ for simplicity and the time-independent Lindblad operators as $L_1=\sqrt{\gamma_1}\,\sigma_-$ and $L_2=\sqrt{\gamma_2}\,\sigma_x$, the Lindblad master equation yields the following differential equations for the Bloch vector components:
\begin{align}
\dot{x}=&-\frac{\gamma_1}{2}x,\notag\\ \dot{y}=&-\left(\frac{\gamma_1}{2}+2\gamma_2\right)y,\notag\\ \dot{z}=&\gamma_1(1-z)-2\gamma_2z.
\end{align}
Solving the above equations gives
\begin{align}
x(t)&=\me^{-\gamma_1 t},\quad y(t)=0,\notag\\ z(t)&=\frac{\gamma_1}{\gamma_1+2\gamma_2}\left(1-\me^{-(\gamma_1+2\gamma_2)t}\right).
\end{align}
The Bures metric components then follow from Eq.~\eqref{A13}:
\begin{align}
g_{rr}(t) = &\frac{1}{4[1 - \me^{-2\gamma_1 t} - \left(\frac{\gamma_1}{\gamma_1 + 2\gamma_2}\right)^2 \left(1 - \me^{-(\gamma_1 + 2\gamma_2)t}\right)^2]},\notag\\
g_{\theta\theta}(t) = &\frac{1}{4}\left[\me^{-2\gamma_1 t} + \left(\frac{\gamma_1}{\gamma_1 + 2\gamma_2}\right)^2(1 - \me^{-(\gamma_1 + 2\gamma_2)t})^2\right],\notag\\
g_{\phi\phi}(t) = &\frac{\me^{-2\gamma_1 t}}{4}.
\end{align}
At $t=0$, we have $r=1$, so the evolution starts from the pure-state boundary. In the limit $t\to0^+$, the radial metric diverges as $g_{rr}\sim 1/t$, but this divergence is integrable, indicating that the system can leave the boundary smoothly. As $t\to\infty$, the Bloch vector ends at $(0,0, \frac{\gamma_1}{\gamma_1+2\gamma_2})$, and the state lies entirely inside the Bloch ball for $0<t<\infty$, corresponding to mixed states. These features are illustrated in Fig.~\ref{Fig3}.

As illustrated in Fig.~\ref{Fig3}(b), the Bloch vector norm $r(t)$ initially decreases from $r=1$ (pure state) as the system enters the interior of the Bloch ball. While the evolution is not purely radial for $t>0$, it is asymptotically radial as $t\to0^+$. The angular motion is of higher order, as confirmed by the analysis below. Correspondingly, Fig.~\ref{Fig3}(c) shows that the radial metric component $g_{rr}$ diverges as $1/t$ when $t\to0^+$; this divergence is integrable, so the Bures distance from the boundary remains finite. Fig.~\ref{Fig3}(d) shows that $g_{\phi\phi}$ decays exponentially (reflecting a loss of phase coherence) while $g_{\theta\theta}$ approaches a constant. This anisotropy indicates that near the Bloch-ball boundary, the dominant contribution to the Bures metric comes from the radial direction. As will be shown below, the scaling $g_{rr}\sim 1/t$ follows from the linear (in time) growth of the smallest eigenvalue of the density matrix.

\subsection{Pure-state Escape Law}\label{Sec:Escape}
Here we present a general property when a system evolves from a pure state according to the Lindblad formalism, which applies to $N\ge 2$. Starting from a pure state $|\psi\rangle$, the projection of the density matrix onto the orthogonal complement $\{|e_a\rangle\}$ of $|\psi\rangle$ as $t\to 0^+$ satisfies
\begin{align}
    \rho_{ab}(t) = t M_{ab} + O(t^2),\quad M_{ab} = \sum_k \langle e_a|L_k|\psi\rangle \langle \psi|L_k^\dagger|e_b\rangle.
\end{align}
Thus the growing eigenvalues of the density matrix away from the pure state are given by the eigenvalues of $M$. Let $\lambda_{(1)}(t)$ denote the largest eigenvalue among these (which controls the overall scale of the growing eigenvalues), and let $\lambda_{\min}(t)$ be the smallest. For $N=2$, they coincide. In general, all eigenvalues of the orthogonal complement block grow linearly:
\begin{align}
    \lambda_{(1)}(t) \sim C_{(1)} t,\qquad \lambda_{\min}(t) \sim C_{\min} t,
\end{align}
with $C_{(1)}=\lambda_{\max}(M)$ and $C_{\min}=\lambda_{\min}(M)$. The radial coordinate in the subsequent conical geometry for $N\ge 3$ is associated with the overall scale factor $\epsilon$ 
, which satisfies $\epsilon(t) \sim C t$ with $C = C_{(1)}$.

When the orthogonal complement is one-dimensional (e.g., for a two-level system), $C_{(1)}=C_{min}=C$ reduces to
\begin{align}\label{C}
    C = \sum_k \bigl( \langle\psi|L_k^\dagger L_k|\psi\rangle - |\langle\psi|L_k|\psi\rangle|^2 \bigr).
\end{align}
For higher-dimensional systems, $C$ is understood as the largest eigenvalue of $M$, which sets the overall scale $\epsilon$ of the conical geometry.

\emph{Proof sketch.} We first expand $\rho(t)$ to first order in $t$ using the Lindblad equation $\dot\rho=\sum_{k}\bigl(L_k\rho L_k^\dagger-\tfrac12\{L_k^\dagger L_k,\rho\}\bigr)$
with $H=0$ for simplicity. It follows that $\rho(t) = |\psi\rangle\langle\psi| + tA + O(t^2)$ with
\begin{align}
    A = \sum_k \left( L_k|\psi\rangle\langle\psi|L_k^\dagger - \frac12 L_k^\dagger L_k|\psi\rangle\langle\psi| - \frac12 |\psi\rangle\langle\psi|L_k^\dagger L_k \right).
\end{align}
We choose an orthonormal basis $\{|\psi\rangle,|e_a\rangle\}$ ($a=1,\dots,N-1$) and obtain 
\begin{align}
    \rho(t) = \begin{pmatrix}
        1 + tA_{00} & tA_{0a} \\
        tA_{a0} & tA_{ab}
    \end{pmatrix} + O(t^2).
\end{align}
The largest eigenvalue of the full matrix is $1+O(t)$ 
while the eigenvalues in the orthogonal complement to $|\psi\rangle$ are $t$ times the eigenvalues of the matrix $M_{ab}=A_{ab}$. Explicitly,
\begin{align}
    A_{ab} = \langle e_a|A|e_b\rangle = \sum_k \langle e_a|L_k|\psi\rangle\langle\psi|L_k^\dagger|e_b\rangle.
\end{align}
The eigenvalues of $M$ (denoted by $c_i$) give the growth rates of the corresponding eigenvalues in the orthogonal space: $\lambda_i(t) \sim c_i t$. In particular, the largest (smallest) eigenvalue of $M$ determine the growth rate of the largest (smallest) among the eigenvalues in the orthogonal complement. 

Applying the above result to the two-level example from pure to mixed states studied previously and substituting the linear behavior of the small eigenvalue into the expression for the radial metric component, $g_{rr}=1/[4(1-r^2)]$ with $r=1-2\lambda_{(1)}$, directly yields $g_{rr}\sim 1/t$, which explains the divergence observed in Fig.~\ref{Fig3}(c).  

For $N\ge3$, the same linear growth holds for all eigenvalues of the orthogonal complement. Within the restricted submanifold where the pure-state direction and the spectral ratios are fixed (see Sec.~\ref{appc}), the Bures metric takes a conical form (see the discussions later). The radial coordinate $u \sim \sqrt{\epsilon}$ (with $\epsilon$ the overall scale factor of the eigenvalues in the orthogonal complement)
scales as $\sqrt{t}$, while angular motion is weighted by a factor of order $\epsilon$ and is therefore negligible as the system approaches the pure state of interest. Consequently, any such evolution emerging from (or approaching) a pure state is asymptotically radial, with frozen angular degrees of freedom.

\subsection{Geometric Classification of $N=2$ Lindblad Dynamics}
The three processes for $N=2$ discussed above exhibit qualitatively different behaviors of the only eigenvalue $\lambda_{\min}(t)$ in the orthogonal complement, which directly controls the radial motion (i.e., the distance to the pure-state boundary) and therefore governs the geometric properties on the Bures manifold.

\emph{Asymptotic purification} corresponds to $\lambda_{\min}(t)\propto\me^{-\gamma t}$. The evolution remains strictly inside the full-rank manifold, approaching the pure-state boundary only when $t\to\infty$ along the radial direction (the direction of $\lambda_{\min}$). The process follows a Bures geodesic as the geometrically shortest path.

\emph{Finite-time purification} is characterized by $\lambda_{\min}(t)\sim (T-t)^\alpha$, with the evolution hitting the pure-state boundary at a finite time. For a two-level system, the boundary is smooth. Therefore, the radial metric divergence is a coordinate artifact removable by a coordinate transformation,
and the angular degrees of freedom remain finite. Despite being purely radial, this trajectory is not a Bures geodesic, in contrast to the asymptotic purification case.

\emph{Decoherence from a pure state} describes evolution that escapes from the pure-state boundary along the radial direction (asymptotically as $t\to0^+$), moving into the interior of the mixed-state manifold. The smallest eigenvalue grows linearly, $\lambda_{\min}(t)\sim C t$, where $C$ is given by Eq.~(\ref{C}). This linear growth signals a smooth departure from the pure-state boundary.

This classification shows that in all three processes, the Lindblad evolution near a rank-changing point proceeds radially (i.e., along the direction of the smallest eigenvalue of the density matrix). In two-level systems, singularities of the Bures metric (such as the divergence of $g_{rr}$) at rank-changing points arise solely due to the coordinate choice since no intrinsic curvature singularity occurs (see Appendix~\ref{app5} for the curvature of the Bures metric for $N=2$ systems).

\section{Rank Changing for $N\ge3$ Systems}\label{Sec.4}

\begin{figure}[t]
\centering
\includegraphics[width=2.4in]{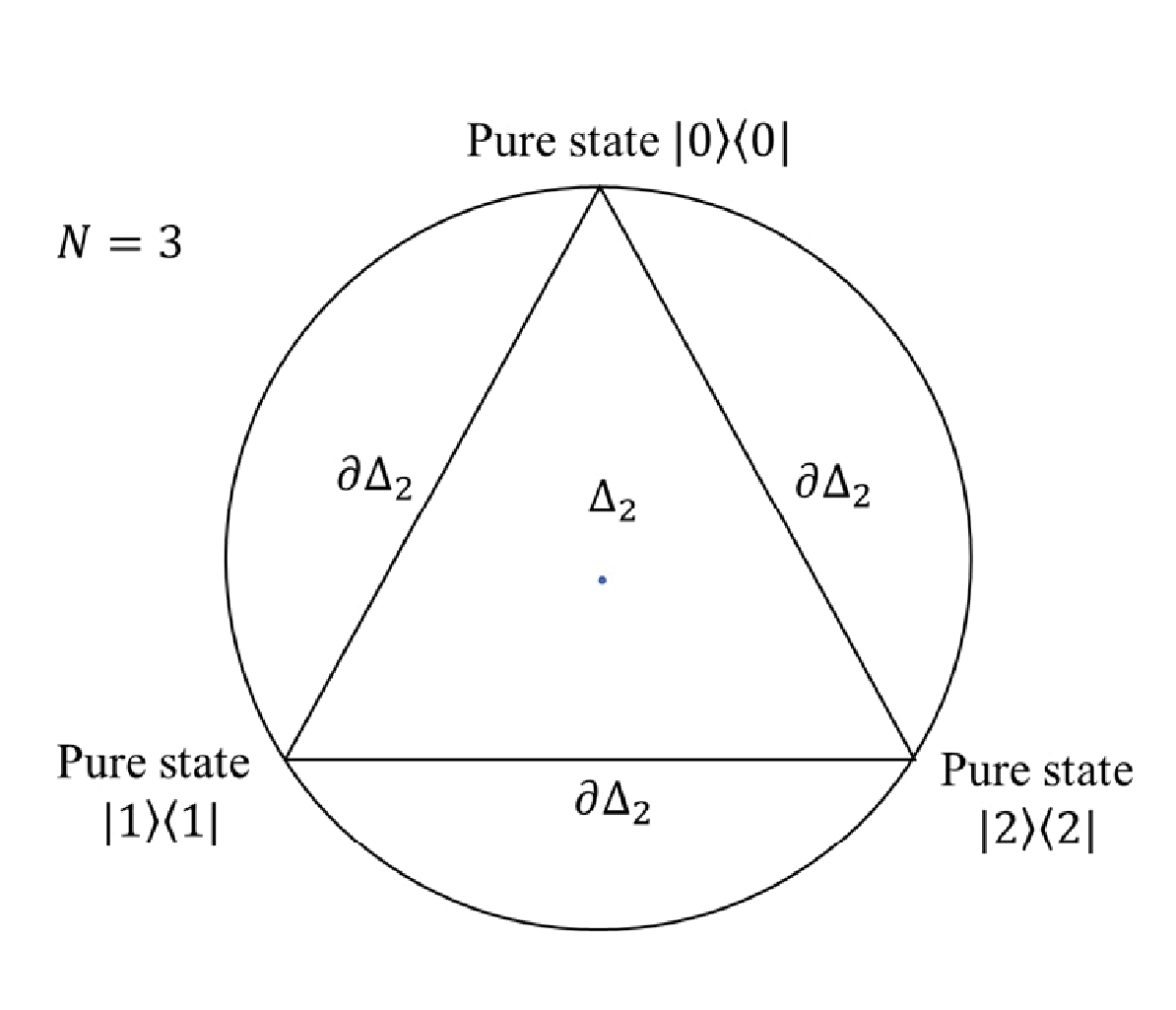}
\caption{Illustration of the geometry of $N=3$ states. The full state space is an $8$-dimensional convex body, represented schematically as a ball. The equilateral triangle $\Delta_2$ is the eigenvalue simplex of diagonal density matrices (Eq.~(\ref{d2})). Its corners are pure states (rank $1$), its edges are rank-$2$ states, and its interior is full rank. Any density matrix is obtained by a unitary transformation of a point in $\Delta_2$. The simplex lies strictly inside the convex body; its edges do not lie on the boundary of the state space. The corners are points where the rank jumps from $3$ to $1$. Near these corners, the geometry develops a conical singularity.
}
\label{N=3}
\end{figure}

\subsection{Stratified Structure and Local Geometry Near Rank-Changing Points}
For a two-level system ($N=2$), the mixed-state manifold is a three-dimensional ball (the Bloch ball) whose boundary is the two-dimensional sphere of pure states (the Bloch sphere). The boundary is smooth, and the Bures metric remains regular when expressed in appropriate coordinates. Consequently, no intrinsic curvature singularity appears at the pure-state boundary when the rank of the density matrix changes.

For $N\ge3$, the geometry of the state space departs radically from the simplicity of the Bloch ball. To understand the origin of this difference, we examine the simplest case $N=3$ (a qutrit or a three-level system). The space of all density matrices for $N=3$ is an $8$-dimensional convex body ($3^2-1=8$). In Fig.~\ref{N=3} we represent this convex body schematically as a ball, but it is important to stress that the actual shape is not a round ball; the drawing merely conveys the idea of a convex set. Inside this body lies the eigenvalue simplex~\cite{Bengtsson_book}
\begin{align}\label{d2}
\Delta_2 = \left\{ \sum_{i=0}^2 \lambda_i |i\rangle\langle i| \;\middle|\; \sum_{i=0}^2 \lambda_i = 1,\; \lambda_i \ge 0 \right\},
\end{align}
which consists of all density matrices that are diagonal in a fixed orthonormal basis $\{|0\rangle,|1\rangle,|2\rangle\}$. The simplex $\Delta_2$ is an analogue of an equilateral triangle. Its three corners correspond to the pure states $|0\rangle\langle0|$, $|1\rangle\langle1|$, and $|2\rangle\langle2|$ (each of rank $1$); its edges correspond to rank-$2$ states (e.g., mixtures of two basis states); and its interior corresponds to full-rank states (all $\lambda_i>0$). Any density matrix can be obtained by applying a suitable unitary transformation to some point in $\Delta_2$; therefore the entire $8$-dimensional convex body is the union of all unitary copies (orbits) of $\Delta_2$. Just as the Bloch ball is filled by rotating a diameter (the $N=2$ eigenvalue simplex) around the origin, the eight-dimensional qutrit state space is generated by sweeping the triangular simplex through all possible unitary rotations.
Each rotated copy of the triangle has its own three vertices, which are the pure states obtained by rotating $|0\rangle\langle0|$, $|1\rangle\langle1|$, $|2\rangle\langle2|$ with the same unitary transformation. Together, the vertices of all such copies constitute the entire pure-state manifold $\mathbb{CP}^2$.

A crucial geometric fact is that the simplex $\Delta_2$ lies inside the convex body, not on its boundary. The edges of $\Delta_2$ are not part of the boundary of the full state space; they are contained in the interior. The actual boundary of the density matrix space is $7$-dimensional. The set of pure states, which for $N=3$ is the complex projective plane $\mathbb{CP}^2$, is only a $4$-dimensional subset of this $7$-dimensional boundary. In other words, for $N\ge3$ the pure-state manifold does not constitute the entire boundary of the mixed-state space. Rather, it is a proper subset of the boundary, and the rest of the boundary consists of mixed states of rank $2$ (and, for higher $N$, of various intermediate ranks). This is in stark contrast to the $N=2$ case, where the pure states (the Bloch sphere) exactly coincide with the boundary of the mixed states (the Bloch ball). Thus, pure states are sparse within the boundary for $N\ge 3$, and the boundary itself is a mosaic of different-dimensional strata. For the $N=3$ case, the edges (rank-2 states) of the eigenvalue simplex are three-dimensional submanifolds embedded in the seven-dimensional boundary, and the vertices (pure states) are zero-dimensional points embedded within those edges.

Now consider a corner of the triangle (eigenvalue simplex), say the pure state $|1\rangle\langle1|$. At this point the density matrix has rank $1$. Moving slightly away from this corner into the interior, the smallest eigenvalue becomes positive and the null subspace, which is the eigenspace of the zero eigenvalue, acquires dimension $2$, carrying internal angular degrees of freedom (rotations within the subspace). A change of the rank from $3$ to $1$ is accompanied by the appearance of those angular directions, which prevent the boundary from being smooth. The local geometry resembles a conical structure: A corner of the triangle is the cone tip. The two edges emanating from the tip represent the directions where the density matrix involves only two basis states (e.g., mixtures of $|2\rangle$ and $|3\rangle$), hence its rank is at most $2$. The interior of the triangle near the corner corresponds to full-rank states. Unitary transformations sweeping out the neighborhood generate a conical shape. In the full $8$-dimensional space for $N=3$, the radial direction (corresponding to the overall scale of the eigenvalues of the orthogonal complement to the tip) will result in curvature singularities, while the angular directions (rotations within the null subspace) are smooth but become frozen at the tip with their metric contributions suppressed. This conical structure is absent for $N=2$, where the null subspace is one-dimensional while the pure-state boundary remains smooth.

In the following, we first focus on the simplest nontrivial case with $N=3$ and then generalize to arbitrary $N\ge3$. 
We will make the above intuitive picture precise by restricting to submanifolds where the conical structure becomes manifest and then compute the resulting curvature singularities.

\subsection{Metric Cone}
To understand geometric singularities arise at rank-changing points for $N\ge 3$, we introduce the notion of a metric cone and show how the Bures metric can be reduced to this form under suitable restrictions. 

\subsubsection{Geometric picture of a metric cone}
A metric cone is a Riemannian manifold with a singularity at a point (the cone tip) with a coordinate realization of the conical metric $\dif s^2 = \dif u^2 + u^2 h$ (see, e.g., Ref.~\cite{Petersen16}). The line element takes the explicit form
\begin{align}\label{ds2du2}
\dif s^2 = \dif u^2 + u^2 h_{ab}(\theta)\dif\theta^a\dif\theta^b,
\end{align}
with $u\ge0$ being the radial coordinate which measures the distance from the tip and $h_{ab}(\theta)$ denoting a metric on a compact base manifold $\Sigma$ (i.e., the ``cross-section" or ``floor" of the cone, orthogonal to the radial direction). The coordinates $\theta^a$ ($a=1,\dots,d$, where $d$ is the dimension of $\Sigma$) parametrize the angular directions around the cone axis. When $u=0$, all points on $\Sigma$ collapse to the singular tip, and for $N\ge3$ the scalar curvature will diverge (the precise behavior is discussed later).

To visualize the geometry, one may consider an ordinary ice-cream cone in three-dimensional Euclidean space with the tip at $u=0$ and moving outward along the radial lines (the straight lines emanating from the tip) as $u$ increases. The base $\Sigma$ is the circular rim at fixed $u$ (a one-sphere $S^1$), and the angular coordinate $\theta$ describes positions on this rim. The line element would be $\dif s^2 = \dif u^2 + u^2 \dif\theta^2$ for a flat cone  (i.e., a cone with zero intrinsic curvature, which can be cut and flattened into a plane), or $\dif s^2 = \dif u^2 + u^2 R^2 \dif\theta^2$ if the rim has radius $R$.

In the full density matrix space, the pure-state manifold is not a cone tip since the metric on the pure-state manifold is the smooth Fubini-Study metric on $\mathbb{CP}^{N-1}$ (i.e., the gauge-invariant metric for pure states). However, when restricted to a suitable submanifold (e.g., fixing a pure state $|\psi\rangle$ and the density-matrix spectrum of the orthogonal complement to $|\psi\rangle$), the Bures metric near a pure state takes the form of a metric cone. In that setting the scalar curvature diverges for $N\ge3$ while for $N=2$ the base $\Sigma$ collapses to a point and the cone reduces to a smooth radial line. Thus, the conical singularity appears only when $N\ge3$.

\subsubsection{Reduction of the Bures metric to a conical metric on a restricted submanifold}
We now construct a submanifold of the full density matrix space on which the Bures metric takes the form of a conical metric near a pure state. The construction is specific to the chosen constraint with fixed $|\psi\rangle$ and fixed spectrum, and under the conditions specified below, the Bures metric near a pure state reduces to a conical metric.
Specifically, we consider a density matrix close to a pure state $|\psi\rangle\langle\psi|$ with eigenvalues $\lambda_0 = 1-\epsilon$ and $\lambda_a = \epsilon\mu_a$ ($a=1,\dots,N-1$), where $\epsilon\ll1$ controls the scale of the eigenvalues in the orthogonal complement to $|\psi\rangle$, and $\mu_a>0$ with $\sum_a\mu_a=1$. Here $\mu_a$ are independent of $\epsilon$. Thus, as $\epsilon\to0$, the entire orthogonal complement of $|\psi\rangle$ (spanned by the eigenvectors of the $\lambda_a$) becomes a null subspace of dimension $D=N-1\ge2$ for $N\ge3$.
The vanishing $\epsilon$ corresponds to a rank change of the density matrix from $N$ to $1$, with the small $N-1$ eigenvalues approaching zero simultaneously. 
Moreover, we impose two constraints: The pure-state direction $|\psi\rangle$ is kept fixed, and the spectrum $\{\mu_a\}$ ($a=1,\dots,N-1$) of the projected density matrix $\sigma$ onto the orthogonal complement is held constant. Only the orthonormal basis $\{|a(\theta)\rangle\}$ of the orthogonal complement is allowed to vary through unitary transformations. Physically, this corresponds to rotating the ``axes" of the complement subspace while keeping its ``shape" (eigenvalues) and the pure-state direction fixed.
We caution that the metric cone near a pure state does not apply to the zero-temperature limit of thermal states, where the approaching to a pure state proceeds via the tangential sector (see Appendix~\ref{appC2}).

Under the above restrictions, the Bures metric reduces to the standard conical form shown in Eq.~\eqref{ds2du2} (a proof is given in Appendix~\ref{appc}), where:
\begin{itemize}
    \item $u \sim\sqrt{\epsilon}$ is the radial coordinate measuring the distance from the rank-changing point;
    \item $\theta^a$ are angular coordinates parametrizing the orientation of the orthonormal basis $\{|a\rangle\}$ within the null subspace;
    \item $h_{ab}(\theta)$ is the metric on the base $\Sigma$, which is the space of orthonormal frames of the null subspace. For non-degenerate $\mu_a$, $\Sigma$ is a homogeneous space diffeomorphic to $\mathbb{CP}^{N-2}$ (or a product of such spaces when degeneracy occurs).
\end{itemize}

Thus, in the limit $\epsilon\to0$, the Bures metric induced on this restricted submanifold becomes a metric cone over the base and it metric $(\Sigma,h)$. This construction isolates the conical singularity that appears in the full density matrix space for $N\ge3$ and reveals how the internal angular degrees of freedom of the null subspace produce the conical geometry. We note the base dimension $d\le D$. For $N=2$, the orthogonal complement is one-dimensional and $\Sigma$ reduces to a single point; the cone collapses to a smooth radial line, consistent with the smooth boundary observed earlier. With this geometric picture in hand, we now study the geodesic behavior of Lindblad evolution near the cone tip.

\begin{figure}[t]
\centering
\includegraphics[width=2.2in]{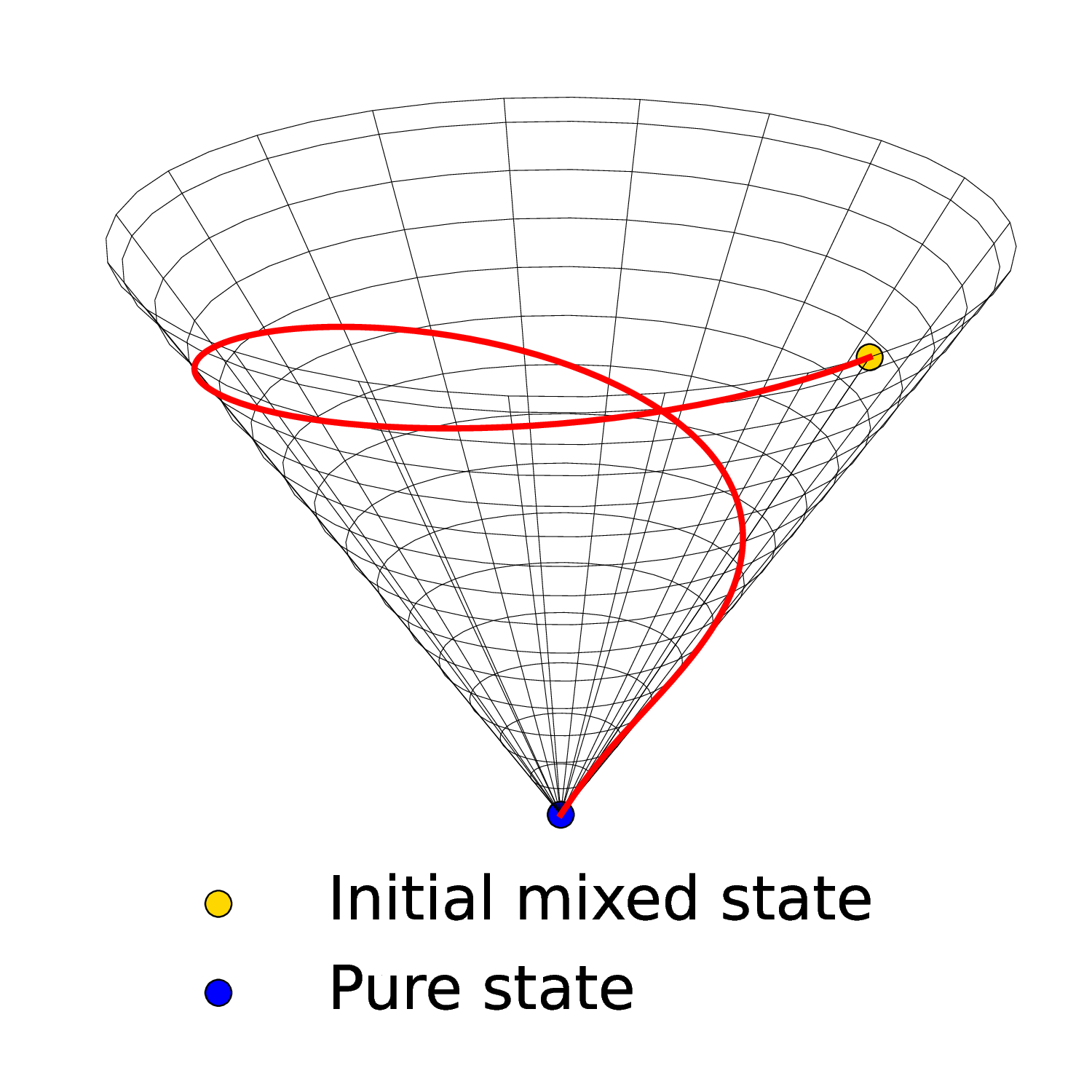}
\caption{Illustration of the metric cone structure near a pure state for $N\ge3$. The yellow dot at the tip is a pure state and also a rank-changing point. The blue dot is a mixed state in the interior. The red curve shows a radial Bures geodesic approaching or shooting from the cone tip, illustrating the ``geodesic approaching/shooting'' behavior of Lindblad evolution. Any evolution that approaches or leaves the tip must become asymptotically radial to avoid divergent angular velocity.}
\label{Fig5}
\end{figure}

\subsection{Geodesic Approaching/Shooting for $N\ge3$}
\subsubsection{Radial constraint from cone geometry}
Our analysis of two-level systems ($N=2$) revealed that near a rank-changing point, all three Lindblad evolutions proceed radially, i.e., along the direction determined by the smaller eigenvalue of the density matrix. Note that in those cases the pure-state boundary is smooth. For $N\ge3$, however, the geometry near a rank-changing point (at the boundary) is fundamentally different: the Bures manifold is locally a metric cone of the form given by Eq.~(\ref{ds2du2}) (see Fig.~\ref{Fig5}), with $u = \sqrt{\epsilon}$ and a genuine conical singularity at $u=0$. Remarkably, the following geodesic analysis shows that a radial constraint emerges as well, forcing all physical evolution to become asymptotically radial. In this setting the geodesic equations (see Appendix~\ref{appd}) become, in the limit $u\to0$,
\begin{align}\label{conegeo}
	\ddot u - u|\dot\theta|^2 = 0,\qquad \frac{\dif}{\dif s}(u^2\dot\theta^a) = 0.
\end{align}
The second equation holds approximately near the cone tip, and the omitted term is discussed in Appendix~\ref{appd}.

Suppose the evolution makes the system approach (or shoot from) the cone tip where $u\to0$. If the conserved quantity $u^2\dot\theta^a$ were nonzero, then $\dot\theta^a = c^a/u^2$ (where $c^a$ is a constant) would diverge, requiring an infinite angular velocity. Substituting into the radial geodesic equation gives $\ddot u \sim 1/u^3$, which would prevent the evolution from reaching $u=0$ smoothly. In order for the system to reach the cone tip at finite arc length, the conserved quantity must vanish, i.e., $\dot\theta^a=0$. Hence the evolution becomes asymptotically radial.

This leads to the following picture: For $N\ge3$, any Lindblad evolution that approaches a pure state (or more generally a rank-changing point) is either shot from or approaches the cone tip along a radial Bures geodesic. The direction is determined by the Lindblad operators, and the segment of the evolution is a radial geodesic connecting the tip and the interior of the state space. 
A detailed proof of the radial geodesic property under the relevant constraints is given in Appendix~\ref{appd}. For a two-level system ($N=2$) the Bures manifold is smooth up to the pure-state boundary (the Bloch sphere). Consequently, no conical singularity appears, and geodesics can reach the boundary from any direction without the radial constraint described above. The cone picture and the associated radial approaching/shooting are therefore features that arise only when the null subspace has dimension higher than $2$, i.e., for $N\ge3$ systems.

\subsubsection{An explicit Lindblad example for $N=3$}
To illustrate the geometric picture, we consider a simple $N=3$ Lindblad model with the following jump operators
\begin{align}
L_1 = \sqrt{\gamma}\,|1\rangle\langle 0|,\qquad L_2 = \sqrt{\gamma}\,|2\rangle\langle 0|,
\end{align}
and set $H=0$ for simplicity. The initial state is the pure state $|\psi\rangle = |0\rangle$, i.e., $\rho(0)=|0\rangle\langle0|$. Physically, these Lindblad operators describe the evolution of the initial pure state into the two orthogonal directions, i.e., shooting out of the tip. The orthogonal complement of $|0\rangle$ is spanned by $|1\rangle$ and $|2\rangle$, which is two-dimensional.

To obtain the full time evolution, we solve the Lindblad master equation
\begin{align}
\dot\rho = \sum_{k=1}^2 \left( L_k\rho L_k^\dagger - \frac12\{L_k^\dagger L_k,\rho\} \right).
\end{align}
Because $L_k^\dagger L_k = \gamma |0\rangle\langle0|$ for both $k$, we have $\sum_k L_k^\dagger L_k = 2\gamma |0\rangle\langle0|$. The equation preserves the diagonal form since the terms $L_k\rho L_k^\dagger$ only contribute to the $|1\rangle\langle1|$ and $|2\rangle\langle2|$ entries while the anticommutator does not introduce off-diagonal elements. Therefore, if $\rho(t)$ is diagonal in the basis $\{|0\rangle,|1\rangle,|2\rangle\}$ at $t=0$, it remains diagonal for all $t$. Writing $\rho(t)=\operatorname{diag}(\rho_{00},\rho_{11},\rho_{22})$, the equations reduce to
\begin{align}
\dot\rho_{00} = -2\gamma\rho_{00},\quad
\dot\rho_{11} = \gamma\rho_{00},\quad
\dot\rho_{22} = \gamma\rho_{00},
\end{align}
with $\rho_{00}+\rho_{11}+\rho_{22}=1$. Solving the equations with $\rho_{00}(0)=1$ and $\rho_{11}(0)=\rho_{22}(0)=0$ gives
\begin{align}
\rho_{00}(t) = \me^{-2\gamma t},\quad
\rho_{11}(t) = \rho_{22}(t) = \frac12\bigl(1-\me^{-2\gamma t}\bigr).
\end{align}
Thus the eigenvalues are $\lambda_0=\me^{-2\gamma t}$, $\lambda_1=\lambda_2=\frac12(1-\me^{-2\gamma t})$. For short times, both eigenvalues in the orthogonal complement satisfy $\lambda_1=\lambda_2\sim \gamma t$, confirming the linear growth and the pure-state escape law (see Sec.~\ref{Sec:Escape}). The $t\rightarrow 0^+$ scaling is precisely the radial behavior required for a metric cone, with $\epsilon \sim \gamma t$ playing the role of the radial parameter. 

The two-dimensional orthogonal complement supports internal angular degrees of freedom. Consequently, the asymptotic Bures metric takes the form of a three-dimensional cone whose base manifold $\Sigma$ describes the orientation of the two-dimensional null subspace. In the above diagonal example, the basis $\{|1\rangle,|2\rangle\}$ is fixed, so $\dot\theta^a=0$ holds trivially. The general argument in Appendix~\ref{appd} ensures that for any physical Lindblad process approaching the cone tip, the radial constraint $\dot\theta^a=0$ must hold, even when angular degrees of freedom are present. Consequently, the evolution must be asymptotically radial and follows a geodesic, as explained in the previous subsection and illustrated by the radial geodesic in Fig.~\ref{Fig5} near the cone tip. This model thus provides an explicit, analytically solvable realization of the geodesic approaching/shooting picture for $N=3$ and may be generalized to higher values of $N$.


\subsection{Curvature Singularities for $N\ge 3$}
In a preceding subsection we showed that under appropriate restrictions (a fixed pure state $|\psi\rangle$ and fixed spectrum $\{\mu_a\}$ of the density matrix), the Bures metric reduces to a metric cone near a rank-changing point. We now investigate the curvature of such a conical geometry.
For two-level systems the pure-state boundary is smooth and the curvature from the Bures metric, $R$, remains constant ($R=24$, as shown in Appendix~\ref{app5}). For $N\ge3$, the null subspace has dimension $D\ge2$, which gives rise to a genuine conical singularity. The scalar curvature typically diverges, but the behavior depends on angular degrees of freedom. Below we illustrate these phenomena with two $N=3$ models, serving as explicit examples of the formal curvature formulas derived in Appendix~\ref{appe}.

Before diving into the singular behavior around rank-changing points, however, it is worth noting that in the zero-temperature limit when $T\to0$, the Bures metric of mixed states in thermal equilibrium reduces to the corresponding Fubini-Study metric for pure states~\cite{PhysRevB.110.035144}. The Fubini-Study metric typically does not exhibit singular behavior near rank-1 pure-state points. Nevertheless, there is no contradiction because, as explained in Appendix~\ref{appC2}, the zero-temperature limit approaches the rank-1 pure-state point through the tangential sector in its neighborhood, whereas the conic construction here probes the transverse sector responsible for the conical singularities. As elaborated in Appendix~\ref{appC2}, the two different ways of approaching a pure state depend on the eigen-spectrum of the density matrix.

\subsubsection{A reduced $N=3$ model with one angular coordinate}\label{Sec2DCone}
We first consider a reduced $N=3$ example where only a single angular degree of freedom near the pure-state cone tip is retained. Let the eigenvalue spectrum of the density matrix be 
\begin{align}\label{evr}
\lambda_0=\epsilon,\quad
\lambda_1=\zeta\epsilon,\quad
\lambda_2=1-(1+\zeta)\epsilon,
\end{align}
with $0<\zeta<1$ being a constant and $\epsilon\to0^+$.
As $\epsilon\to0$, the density matrix tends to the pure state $|2\rangle\langle2|$ and the rank drops from $3$ to $1$. The two small eigenvalues scale with the same power but maintain a fixed ratio $\zeta$.

 \begin{figure}[t]
\centering
\includegraphics[width=3.2in]{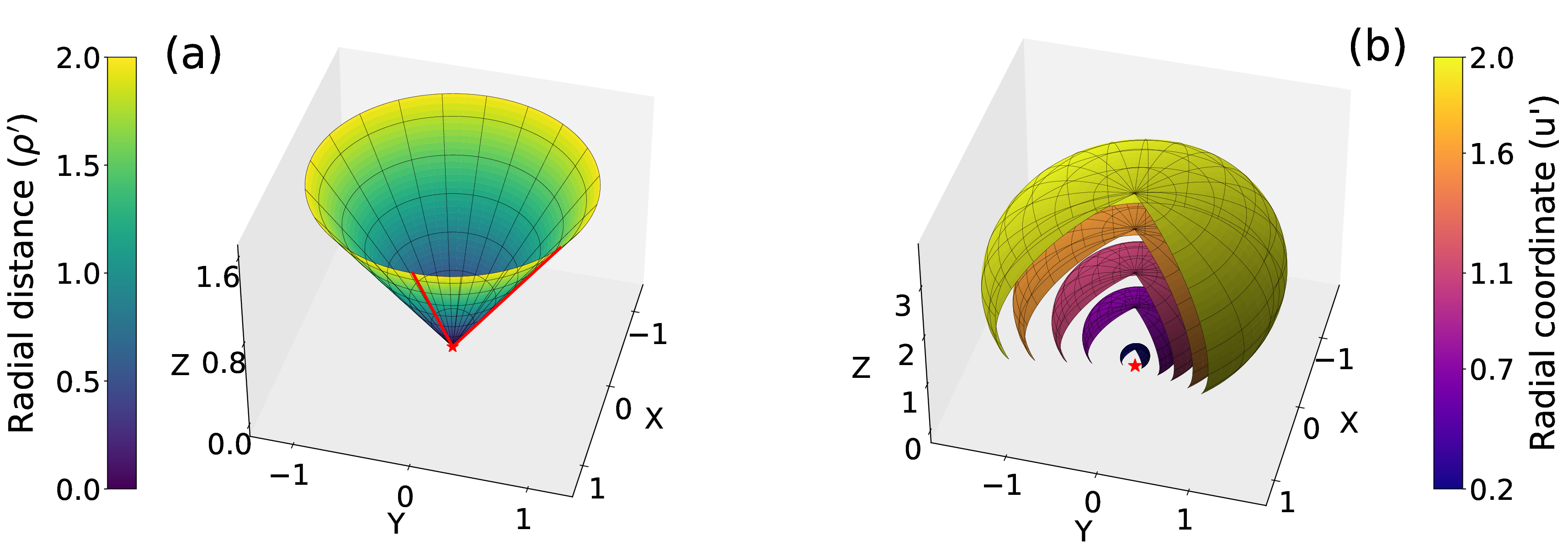}
\caption{(a) A 2D conical singularity described by Eq.~(\ref{cone_metric2D}) with $\kappa=0.6$. The axes $X,Y,Z$ denote the Cartesian coordinates of the cone embedded in $\mathbb{R}^3$, dependent on the radial parameter and the angular coordinate. (b) Cutaway view of the 3D metric cone of Eq.~(\ref{cone_metric3D}) with $\kappa=0.8$, showing nested spherical cross-sections at constant $u'$.}
\label{Fig6}
\end{figure}

To obtain the simplest angular dependence, we introduce a single-parameter rotation $U(\theta)=\me^{-\mi\theta G}$ between $|0\rangle$ and $|1\rangle$ with the generator
\begin{align}\label{G2D}
G=\frac{\mi}{2}\bigl(|0\rangle\langle 1|-|1\rangle\langle 0|\bigr).
\end{align}
The eigenstates become $|i(\theta)\rangle=U(\theta)|i\rangle$; the only non-zero angular inner product is
\begin{align}
\langle 0|\dif 1\rangle=\frac{\dif\theta}{2},\quad \langle 1|\dif 0\rangle=-\frac{\dif\theta}{2},
\end{align}
where we have suppressed the explicit $\theta$-dependence of the rotated states for brevity. 
Since only one angular degree of freedom is retained, the base manifold $\Sigma$ is a circle ($d=1$), and the resulting conical geometry is two-dimensional (one radial coordinate $u$ plus one angular coordinate $\theta$). This corresponds to the case $d=1$ in the general conical metric of Eq.~\eqref{ds2du2}.

The Bures metric in spectral representation reads
\begin{align}\label{eq:bures_gen}
\dif s^2
=
\frac14\sum_i\frac{(\dif\lambda_i)^2}{\lambda_i}
+
\frac12\sum_{i<j}
\frac{(\lambda_i-\lambda_j)^2}{\lambda_i+\lambda_j}
\,|\langle i|\dif j\rangle|^2.
\end{align}
For the radial part we have
\begin{align}
\dif s_{\mathrm{rad}}^2
=&
\frac14\left(
\frac{1+\zeta}{\epsilon}
+
\frac{(1+\zeta)^2}{1-(1+\zeta)\epsilon}
\right)\dif\epsilon^2.
\end{align}
In the limit $\epsilon\to0$ the dominant term is
\begin{align}\label{eq:rad_asymp}
\dif s_{\mathrm{rad}}^2 \approx \frac{1+\zeta}{4\epsilon}\,\dif\epsilon^2.
\end{align}
Only the $(0,1)$ pair of the states contributes to the angular part, with coefficient
$\frac12\frac{(\lambda_0-\lambda_1)^2}{\lambda_0+\lambda_1}
=
\frac{(1-\zeta)^2}{8(1+\zeta)}\,\epsilon$.
Thus
\begin{align}\label{eq:ang_asymp}
\dif s_{\mathrm{ang}}^2 = \frac{(1-\zeta)^2}{8(1+\zeta)}\,\epsilon\,\dif\theta^2.
\end{align}
Combining the radial and angular contributions gives
\begin{align}\label{eq:asymp_metric}
\dif s^2
\approx
\frac{1+\zeta}{4\epsilon}\,\dif\epsilon^2
+
\frac{(1-\zeta)^2}{8(1+\zeta)}\,\epsilon\,\dif\theta^2.
\end{align}
Defining $u'=\sqrt{(1+\zeta)\epsilon}$ and $\kappa^2=(1-\zeta)^2/[8(1+\zeta)^2]$, we obtain the standard two-dimensional conical metric
\begin{align}\label{cone_metric2D}
\dif s^2=\dif u'^2+\kappa^2 u'^2\dif\theta^2,\quad u'\ge0,\ \theta\in[0,2\pi).
\end{align}
Hence, in this reduced sector, the Bures geometry near the rank-changing point is a two-dimensional cone with a metric of the form of Eq.~\eqref{ds2du2}.

For $u'>0$ the geometry is locally flat, and when $\kappa\neq1$ there is an angular defect at the tip $u'=0$. The scalar curvature $R$ from the Bures metric is related to the Gaussian curvature $K$ via $R=2K$~\cite{do_carmo_book}. However, $K$ cannot be obtained from the standard local Riemann tensor, which vanishes for $u'>0$. Instead, we show that it is concentrated at the cone tip. By applying the Gauss-Bonnet theorem for orbifolds (see, e.g., Ref.~\cite{Seaton2008}) to a small disk surrounding the tip, one deduces
\begin{align}
K = (1-\kappa)\,\frac{\delta(u')}{u'},
\end{align}
where $\delta(\rho')$ is the one-dimensional Dirac delta function.  Therefore,
\begin{align}\label{Eq:R_Delta}
R = 2(1-\kappa)\,\frac{\delta(u')}{u'}.
\end{align}
Alternatively, using the two-dimensional area element
$\dif A=u'\,\dif u'\dif\theta$,
the integrated Gaussian curvature over a small disk surrounding the tip is
\begin{align}
\int_D K\,\dif A
=
2\pi(1-\kappa),
\end{align}
which is precisely the angular deficit of the cone~\cite{Seaton2008}.

\begin{figure}[t]
\centering
\includegraphics[width=3.2in]{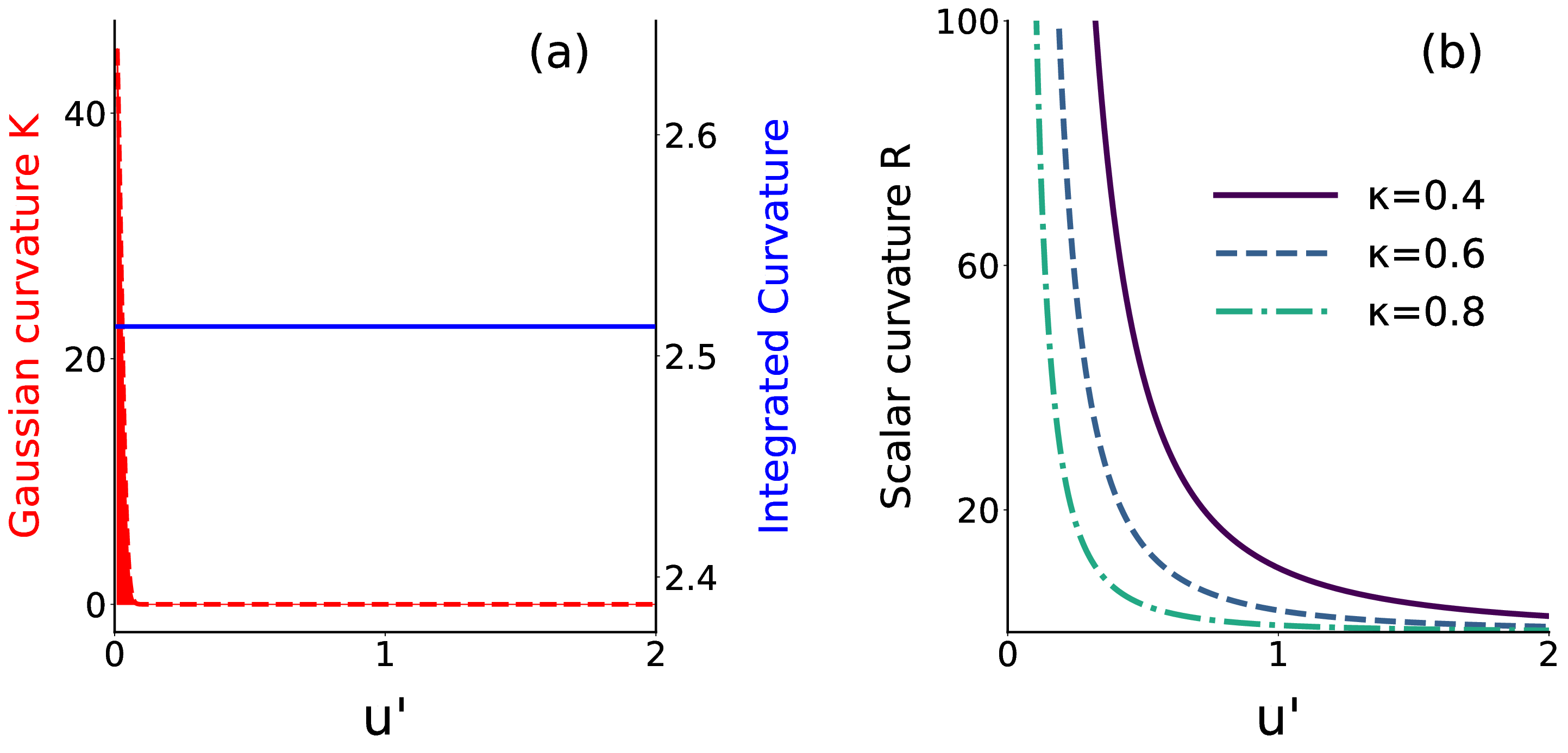}
\caption{(a) Gaussian curvature $K$ (red dashed curve) of the 2D cone with $\kappa=0.6$, concentrated as a Dirac delta-function at the tip. The blue solid line shows the integrated deficit angle $\delta=2\pi(1-\kappa)$. (b) Scalar curvature $R$ of the 3D cone as a function of $u'$ for selected values of $\kappa$, displaying a $1/u'^2$ divergence as $u'\to0$.
}
\label{Fig7}
\end{figure}

Fig.~\ref{Fig6}(a) and Fig.~\ref{Fig7}(a) illustrate the geometry of the 2D conical singularity. Fig.~\ref{Fig6}(a) shows the cone embedded in $\mathbb{R}^3$. The tip (red star) carries an angular deficit $2\pi(1-\kappa)=2.51\ \mathrm{rad}$ for $\kappa=0.6$.  Fig.~\ref{Fig7}(a) shows the corresponding Gaussian curvature $K$ (red dashed curve), which vanishes for $u'>0$ and is concentrated at the cone tip, yielding the constant, integrated curvature.
Unlike the Gauss-Bonnet relation for smooth closed manifolds, the integrated curvature of the metric cone is generally not quantized because the cone tip is an orbifold singularity rather than a regular manifold point~\cite{Seaton2008}. The angular-defect strength is continuously controlled by the spectral ratio $\zeta$, so the curvature concentrated at the rank-changing point can take non-integer values. The singularity should therefore be understood as a geometric feature of the mixed-state manifold, leading to nontrivial geometric properties for the geodesic and curvature near the cone tip.

\subsubsection{Full angular structure for $N=3$: A three-dimensional metric cone}
We present a more general example with a full metric cone for the $N=3$ system. For the radial part of the Bures metric, i.e., the contribution from variations of the eigenvalues, we use the same forms as those in the reduced $N=3$ model given by Eq.~(\ref{evr}). In the previous subsection we retained only a single angular degree of freedom ($\theta$) of the null space, which leads to a two-dimensional cone. Here we enlarge the angular freedom to its full dimension by allowing rotations that span the entire two-dimensional null subspace (isomorphic to $\mathbb{CP}^1\cong S^2$). As a result, the base manifold $\Sigma$ becomes two-dimensional, producing a genuine three-dimensional metric cone.

The unitary transformation now depends on two angles $(\theta,\phi)$. We generalize the generator in Eq.~(\ref{G2D}) to
\begin{align}\label{U3DCone}
G = \frac{\mi}{2}\bigl(\me^{\mi\phi}|0\rangle\langle1| - \me^{-\mi\phi}|1\rangle\langle0|\bigr),
\end{align}
and set $U(\theta,\phi) = \me^{-\mi\theta G}$. After the rotation, the eigenstates of $\rho$ become
\begin{align}
|0(\theta,\phi)\rangle &= \cos\frac{\theta}{2}|0\rangle - \me^{-\mi\phi}\sin\frac{\theta}{2}|1\rangle,\notag\\
|1(\theta,\phi)\rangle &= \me^{\mi\phi}\sin\frac{\theta}{2}|0\rangle + \cos\frac{\theta}{2}|1\rangle,
\end{align}
while $|2\rangle$ remains unchanged. (Unlike in the two-level system, here $\theta$ and $\phi$ parametrize the base manifold $\Sigma$ of the cone, not the direction of a single Bloch vector.) A direct calculation gives
\begin{align}
|\langle0|\dif1\rangle|^2 = \frac14(\dif\theta^2+\sin^2\theta\dif\phi^2),
\end{align}
which is $1/4$ times the Fubini-Study metric on $\mathbb{CP}^1$ (see Eq.~(\ref{dFS}) or Ref.~\cite{Bengtsson_book}), and we have suppressed the explicit $(\theta,\phi)$-dependence of the rotated states for brevity. Moreover, orthogonality conditions imply
\begin{align}
\langle 0|\dif 2\rangle = -\langle 2|\dif 0\rangle = 0,\qquad
\langle 1|\dif 2\rangle = -\langle 2|\dif 1\rangle = 0,
\end{align}
so the $(0,2)$ and $(1,2)$ pairs do not contribute to the angular part at leading order. Their contributions are constants (independent of $\epsilon$) and correspond to flat directions that do not affect the conical singularity.

With the same eigenvalue-spectrum as Eq.~(\ref{evr}), the radial part $\dif s_{\mathrm{rad}}^2$ is identical to the previous case described by Eq.~(\ref{eq:rad_asymp}),
For the angular part, only the $(0,1)$ pair contributes a term proportional to $\epsilon$:
\begin{align}\label{eq:ang_full}
\dif s_{\mathrm{ang}}^2 \approx \frac{(1-\zeta)^2}{8(1+\zeta)}\,\epsilon\,(\dif\theta^2+\sin^2\theta\dif\phi^2).
\end{align}
Thus the asymptotic Bures metric in the limit $\epsilon\to0$ is $\dif s^2=\dif s^2_\text{rad}+\dif s^2_\text{ang}$.
Similarly to the 2D cone, we set $u'=\sqrt{(1+\zeta)\epsilon}$ and $\kappa^2 = \frac{(1-\zeta)^2}{8(1+\zeta)^2}$ and obtain
\begin{align}\label{cone_metric3D}
\dif s^2 = \dif u'^2 + \kappa^2 u'^2(\dif\theta^2+\sin^2\theta\dif\phi^2).
\end{align}
This is a three-dimensional metric cone ($d=2$), in contrast to the two-dimensional cone ($d=1$) obtained in the reduced model where only the $\theta$ coordinate in the null space was kept. We note that setting $\phi=0$ reduces Eq.~(\ref{cone_metric3D}) to the two-dimensional conic metric of Eq.~(\ref{cone_metric2D}), but the curvature singularities of the two cases are qualitatively different. 
The presence of the additional angular coordinate $\phi$ (resulting in the full $S^2$ base) changes the nature of the singularity: The scalar curvature associated with the Bures metric now scales as $R\sim u'^{-2}$, a power-law divergence, rather than the delta-function concentration of the reduced model with only one null-space coordinate. This distinction highlights how the dimensionality of the base manifold governs the type of curvature singularity at rank-changing points.



 \begin{figure}[t]
\centering
\includegraphics[width=3.2in]{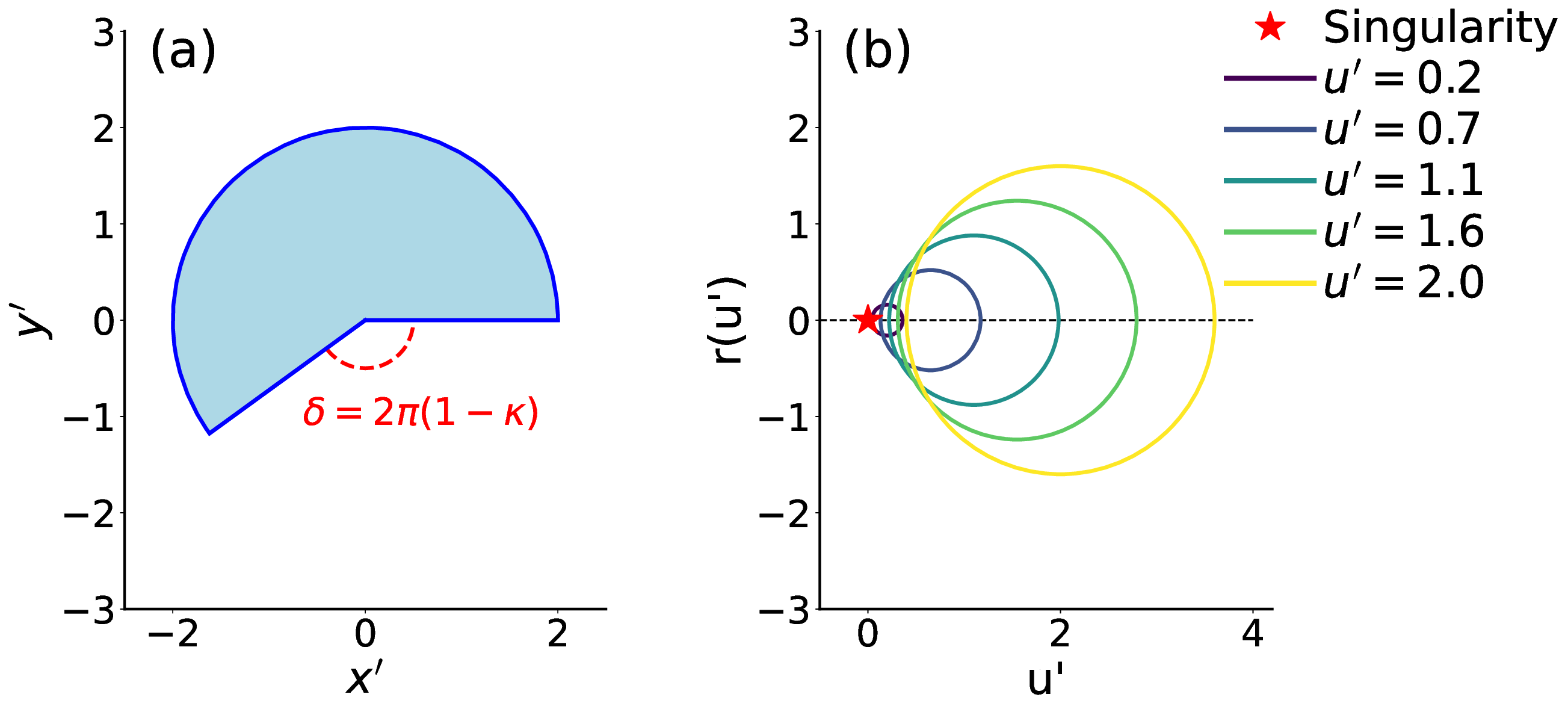}
\caption{(a) Unfolded view of the 2D cone as a planar sector with angle $2\pi\kappa$, showing the angular deficit $\delta=2\pi(1-\kappa)$ (red dashed arc). (b) Cross-section of the 3D metric cone displaying nested spherical cross-sections of radius $r=\kappa u'$ at various radial coordinates $u'$ (color-coded), all converging to the conical singularity at $u'=0$ (red star).
}
\label{Fig8}
\end{figure}

For a conical metric $\dif s^2 = \dif u'^2 + u'^2 h$~\cite{Petersen16}, 
$R = (R_h-2)/u'^2$,
where $R_h$ is the scalar curvature of $h$. Here $h = \kappa^2(\dif\theta^2+\sin^2\theta\dif\phi^2)$ is the metric of a sphere of radius $\kappa$, so $R_h = 2/\kappa^2 = 16(1+\zeta)^2/(1-\zeta)^2$. Hence (see Appendix~\ref{appe})
\begin{align}
R = \frac{ \dfrac{16(1+\zeta)^2}{(1-\zeta)^2} - 2 }{u'^2}
   = \frac{2\bigl[8(1+\zeta)^2-(1-\zeta)^2\bigr]}{(1-\zeta)^2\,u'^2}.
\end{align}
For $\zeta\neq1$, the scalar curvature diverges as $1/u'^2$ when $u'\to0$, signaling a genuine intrinsic curvature singularity at the rank-changing point. (When $\zeta=1$ the two small eigenvalues are equal; then the contraction factor becomes zero at leading order, but the full geometry still exhibits a three-dimensional metric cone with the same type of divergence, as can be verified by a separate analysis.)

Similarly, Fig.~\ref{Fig6}(b) and Fig.~\ref{Fig7}(b) illustrate the geometry of the 3D conical singularity. Fig.~\ref{Fig6}(b) displays a cutaway view of the 3D metric cone, where successive spherical cross-sections at increasing $u'$ reveal the conical structure. Fig.~\ref{Fig7}(b) displays the scalar curvature $R$ of the 3D cone as a function of the radial coordinate $u'$ for several values of $\kappa$. All curves diverge as $1/u'^2$ when $u'\to0$, confirming the power-law singularity predicted by the general theory. Moreover the magnitude of the divergence increases as $\kappa$ decreases (i.e., as the angular deficit grows). Together, these panels illustrate the two distinct types of curvature singularities in the Bures geometry of higher-dimensional systems. The power-law divergence $R\sim u'^{-2}$ (for base dimension $d\ge2$) signals a genuine conical singularity featuring growing curvature without bound as the state approaches the pure state and shrinking angular degrees of freedom in the null subspace. 
The concrete example above illustrates that the conical structure derived in the previous subsection indeed produces power-law divergences in the scalar curvature, confirming the formal expressions of Appendix~\ref{appe}.
For $d=1$ (e.g., the reduced model with only one angular coordinate), the curvature instead concentrates as a Dirac delta function (see Eq.~\eqref{Eq:R_Delta}). Thus, the dimensionality of the cone base determines the severity of the curvature singularity.

Furthermore, Figure \ref{Fig8} provides complementary planar visualizations of the metric cone: Panel (a) shows the 2D cone unfolded into a sector, making the angular deficit manifest as the missing wedge. Panel (b) displays a cross-sectional view of the 3D cone where nested circles represent the shrinking spherical cross-sections as $u'\to0$.

\subsubsection{Constructing Lindblad evolution for cone singularities}
An explicit Lindblad construction that can realize the prescribed density matrix family featuring the aforementioned cone singularities proceeds in two steps. For given angles $(\theta,\phi)$, one first applies the unitary transformation $U(\theta,\phi)$ defined in Eq.~(\ref{U3DCone}) to rotate the reference basis $\{|0\rangle,|1\rangle,|2\rangle\}$ into the rotated basis $\{|0(\theta,\phi)\rangle,|1(\theta,\phi)\rangle,|2\rangle\}$. Second, in this rotated basis define four time-independent Lindblad operators
\begin{align}
L_{20}&=\sqrt{\Gamma}\,|2\rangle\langle 0(\theta,\phi)|,\quad
L_{21}=\sqrt{\Gamma}\,|2\rangle\langle 1(\theta,\phi)|,\notag\\
L_{02}&=\sqrt{\Gamma\frac{\epsilon}{1-(1+\zeta)\epsilon}}\;|0(\theta,\phi)\rangle\langle 2|, \notag\\
L_{12}&=\sqrt{\Gamma\frac{\zeta\epsilon}{1-(1+\zeta)\epsilon}}\;|1(\theta,\phi)\rangle\langle 2|,
\end{align}
with $\Gamma>0$ an overall rate. Setting $H=0$ for simplicity, the Lindblad equation $\dot\rho=\sum_{k}\bigl(L_k\rho L_k^\dagger-\tfrac12\{L_k^\dagger L_k,\rho\}\bigr)$ has a unique steady state independent of the initial condition. For concreteness, one may take $\rho(0)=|2\rangle\langle2|$ without loss of generality. Working in the rotated basis, all off-diagonal elements of $\rho$ decay exponentially to zero because each $L_k$ connects only the indicated basis states and no coherent term ($H=0$) is present to regenerate them.

The rate equations for the populations $p_i=\langle i(\theta,\phi)|\rho|i(\theta,\phi)\rangle$ read
\begin{align}
\dot p_0 &= \Gamma\frac{\epsilon}{1-(1+\zeta)\epsilon}\,p_2 - \Gamma\,p_0,\notag\\
\dot p_1 &= \Gamma\frac{\zeta\epsilon}{1-(1+\zeta)\epsilon}\,p_2 - \Gamma\,p_1,\notag\\
\dot p_2 &= \Gamma(p_0+p_1) - \Gamma\frac{(1+\zeta)\epsilon}{1-(1+\zeta)\epsilon}\,p_2.
\end{align}
One can verify directly that the distribution of Eq.~\eqref{evr} ($p_0 = \epsilon$, $p_1 = \zeta\epsilon$, $p_2 = 1-(1+\zeta)\epsilon$)
satisfies $\dot p_i=0$ for any $\zeta>0$ and $\epsilon\in\bigl(0,\frac{1}{1+\zeta}\bigr)$. Hence the steady state is given by 
\begin{align}
\rho_{\mathrm{ss}}& = \epsilon\,|0(\theta,\phi)\rangle\langle 0(\theta,\phi)| + \zeta\epsilon\,|1(\theta,\phi)\rangle\langle 1(\theta,\phi)|\notag\\& + [1-(1+\zeta)\epsilon]\,|2\rangle\langle 2|.
\end{align}
Thus, the required family of density matrices with the eigen spectrum of Eq.~\eqref{evr} is obtained by first rotating the basis and then letting the system evolve under the dissipative dynamics described above. We remark that setting $\phi=0$ reduces the rotation to a single-parameter family by $U(\theta)$, which recovers the previously analyzed model with a two-dimensional cone and a Dirac delta-function singularity of the curvature. With mathematical rigor, the construction provides an explicit physical mechanism via Lindblad evolution that can realize the parameterized density matrix family exhibiting various cone singularities at rank-changing points. 

\section{Implications}\label{Sec:Implications}
We fist emphasize that for higher-level systems ($N>3$), the same conical geometry and power-law divergence of the curvature, $R\sim u'^{-2}$, persist. However, the base manifold becomes $\mathbb{CP}^{N-2}$, enriching the angular structure while the leading curvature singularity remains unchanged, as indicated in Appendix \ref{appe}. Therefore, it suffices to probe the singular behavior due to the conic metric at rank-changing points in $N=3$ systems although $N>3$ systems add more angular degrees of freedom to the metric cone.

From an experimental perspective, the Bures metric and its curvature can in principle be reconstructed from the density matrix measured via quantum state tomography~\cite{PhysRevLett.90.193601,Paris2004}. However, practical challenges include dense sampling of the parameter space and stable numerical differentiation. Moreover, the curvature involves second-order derivatives and is especially sensitive to noise near a rank-changing point where it diverges. One can mitigate the problem by fitting the predicted scaling laws (e.g., $R \sim u'^{-2}$) from the tomography of the density matrix at multiple locations in the parameter space instead of probing the cone tip of the pure state directly.

A more efficient route bypasses a full tomography by targeting the quantum geometric tensor (QGT)~\cite{PhysRevB.88.064304}, whose real part coincides with the Bures metric. Several complementary protocols have been demonstrated, such as periodically driving a control parameter to excite the system at a rate proportional to the local quantum metric to extract the metric tensor in lattice models and superconducting circuits~\cite{PhysRevB.97.201117,PhysRevLett.122.210401}; coherent Rabi response under a parametric modulation encoding the QGT in the amplitude and phase of resonant oscillations to enable complete measurements of both metric and Berry curvature in NV-center qubit systems~\cite{Yu2020}; a sudden quench to directly access the metric components through the transition probabilities to excited states~\cite{PhysRevB.97.201117,PhysRevLett.122.210401}; and microwave spectroscopy in superconducting Josephson junctions to reveal the QGT of topological Andreev bound states~\cite{PhysRevLett.124.197002}. These dynamical-response methods trade the $O(N^{2})$ overhead of a full tomography for the targeted measurement of the geometric susceptibilities and are suitable for low-dimensional systems like qubits and qutrits. The conical singularities predicted here could be observed by tuning temperature or control parameters across a rank-changing point. Consequently, future experiments with improved parameter resolution may confirm the curvature divergence near a pure state through the $1/u'^{2}$ scaling of $N\ge 3$ systems.

\section{Conclusion}\label{Sec.5}
We have presented an analysis of the Bures metric near rank-changing points, motivated by the observation that nonequilibrium open-system dynamics can vary the rank of density matrices and drive the systems from mixed states to pure states or vice versa. 
Our results reveal a fundamental difference between two-level and higher-dimensional quantum systems near rank-changing points. For $N=2$ systems, the density matrix forms the Bloch ball with the Bloch sphere of pure states as its boundary. Therefore, Lindblad evolution can approach or leave pure states without incurring genuine singularities. 
For $N\ge3$, in stark contrast, the stratified structure of the state space can produce conical singularities at rank-changing points. The null subspace orthogonal to a selected pure state supplies nontrivial angular degrees of freedom with asymptotically radial evolution near the cone tip. The resulting curvature singularities
are intrinsic geometric features rather than coordinate artifacts. Depending on the angular degrees of freedom of the conical metric, the scalar curvature exhibits delta-function or power-law singularities at the cone tip.

Our analyses of the local geometry near rank-changing points unveil the complicated structures of mixed states for revealing genuine geometric singularities. Those findings establish a geometric framework for understanding rank-changing processes in open quantum systems. The singular behaviors confirm that rank changes are not only algebraic transitions of density matrices but also geometric phenomena characterized by singular metric structures and associated quantities. The underlying geometric effects of rank changes may inspire searches of experimentally accessible quantities related to those singular structures in the state space to reveal exciting physics at rank-changing points.


\section{Acknowledgments}
H.G. was supported by H. G. was supported by the Quantum Science and Technology-National Science and Technology Major Project (Grant No. 2021ZD0301904) and the National Natural Science
Foundation of China (Grant No. 12447216). X. Y. H. was supported by the Jiangsu Funding Program for Excellent Postdoctoral Talent (Grant No. 2023ZB611). C.C.C. was supported
by the National Science Foundation under Grant No.
PHY-2310656. We thank Prof. Xin-Sheng Tan for thoughtful discussions.

Yu-Huan Huang and Xu-Yang Hou contributed equally
to this work.
\appendix

\section{Curvature of the Bures metric for two-level systems}\label{app5}
The Bures metric for a two-level system in Bloch coordinates is
\begin{align}
\dif s^2 = \frac14\left( \frac{\dif r^2}{1-r^2} + r^2\dif\Omega^2 \right),\quad
\dif\Omega^2 = \dif\theta^2 + \sin^2\theta\,\dif\phi^2.
\end{align}
Introduce $r = \cos u$ ($0\le u\le \pi/2$). Then $\dif r = -\sin u\,\dif u$ and $1-r^2 = \sin^2 u$, so
\begin{align}
\dif s^2 = \frac14\left( \dif u^2 + \cos^2 u\,\dif\Omega^2 \right).
\end{align}
This is a warped product metric. Let $\dif s^2 = a^2\bigl(\dif u^2 + f(u)^2\dif\Omega^2\bigr)$ with $a=1/2$ and $f(u)=\cos u$.
For a three-dimensional metric $\dif\tilde s^2 = \dif u^2 + f(u)^2\dif\Omega^2$, the scalar curvature is
\begin{align}
\tilde R = -4\frac{f''(u)}{f(u)} + 2\frac{1 - f'(u)^2}{f(u)^2}.
\end{align}
Substituting $f(u)=\cos u$, $f'(u)=-\sin u$, $f''(u)=-\cos u$ gives
$\tilde R = -4\frac{-\cos u}{\cos u} + 2\frac{1-\sin^2 u}{\cos^2 u} = 6$.
The original metric is $\dif s^2 = a^2\dif\tilde s^2$, so the scalar curvature scales as $R = a^{-2}\tilde R$. With $a=1/2$, $a^{-2}=4$, so we obtain
\begin{align}
R = 4 \times 6 = 24.
\end{align}
Thus, the pure-state boundary is smooth, and no intrinsic curvature singularity appears.

\section{Tangential and transverse geometries near rank-change points}
\label{appC2}

\subsection{Tangential versus transverse sectors}
Ref.~\cite{PhysRevB.110.035144} has shown that in the zero-temperature limit, the thermal-equilibrium density matrix becomes the rank-1 ground-state. The Bures metric following the limit reduces to the Fubini-Study metric, which is completely regular on the pure-state manifold. Meanwhile, we show that there is another geometric sector near a rank-changing point when $N\ge 3$, where the Bures metric becomes the conical metric of Eq.~\eqref{ds2du2}.

To see how the two structures are compatible, we consider a thermal state $\rho=\sum_i\lambda_i|i\rangle\langle i|$ with $\lambda_i=\frac{1}{Z}\me^{-\beta E_i}$, where $E_0<E_1<\cdots$ (assuming no degeneracy) and $\beta=1/T$ is the inverse temperature. As $T\to0$, $\lambda_0\to1$ and $\lambda_{i>0}\to0$, so $\rho\to |\psi\rangle\langle\psi|$ with $|\psi\rangle=|0\rangle$; that is, the spectrum beyond the ground state is exponentially suppressed. The subtlety lies on a decomposition. Near a rank-1 point, the Bures geometry contains two distinct sectors: a transverse sector giving rise to the conical metric and a tangential sector giving rise to the Fubini-Study metric. The decomposition is a consequence of the stratified nature of the density matrix space. The distinction becomes clear when we examine the tangent space near a rank-1 point. The pure-state manifold $\mathcal D_1$ (the set of rank-1 density matrices) is naturally embedded as a submanifold in the higher-dimensional full-rank manifold $\mathcal D_N$. Near a rank-1 point, the tangent space of the ambient space decomposes into a direct sum:
\begin{align}
T_\rho\mathcal D = T_\rho^{\parallel}\mathcal D_1 \oplus T_\rho^{\perp}\mathcal D_1,
\end{align}
where the two subspaces are characterized as follows.
\begin{enumerate}
    \item \textbf{Tangential space $T_\rho^{\parallel}\mathcal D_1$}: It consists of variations of the state vector $|\psi\rangle$ that preserve the rank-1 condition, i.e., keep the density matrix on the pure-state manifold. They correspond to changes $|\psi\rangle \to |\psi\rangle + |\dif\psi\rangle$, with normalization preserved and $\dif|\psi\rangle$ orthogonal to $|\psi\rangle$ ($\langle\psi|\dif\psi\rangle=0$). Thus $\dif\rho$ has no $|\psi\rangle\langle\psi|$-component, avoiding any deviation from the eigenvalue $1$, and the state remains in the rank-1 submanifold. Such variations represent intrinsic motions inside the pure-state manifold.

    \item \textbf{Normal space $T_\rho^{\perp}\mathcal D_1$}: It consists of variations that change the spectral structure, pushing the state off $\mathcal D_1$ into the full-rank interior. Examples include increasing the smallest eigenvalue $\lambda_{\min}>0$ or rotating the orthonormal basis $\{|a\rangle\}$ of the orthogonal complement while $|\psi\rangle$ is fixed; such variations increase the rank beyond $1$. These directions are transverse (normal) to the pure-state submanifold in the sense of Riemannian geometry.
\end{enumerate}

Thus, variations of the state vectors of the density matrix are called tangential because they move the density matrix within the pure-state stratum, whereas variations that modify the spectrum of the density matrix are called transverse (or normal) because they connect the rank-1 submanifold to the higher-rank interior.
Explicitly, the Fubini-Study metric describes deformations tangent to the pure-state manifold, and the conical metric instead describes transverse motion away from the pure-state manifold into the full-rank region:
\begin{align}\label{Eq:rho_trans}
\rho = (1-\epsilon)|\psi\rangle\langle\psi| + \epsilon\sigma\to  |\psi\rangle\langle\psi|
\end{align}

Although the thermal-equilibrium state also approaches the rank-1 boundary from the interior of the full density-matrix space, its dominant variations become asymptotically tangent to the pure-state manifold. Indeed, as $\beta\to\infty$, $\lambda_0\to1$ and $\lambda_{n>0}\sim\me^{-\beta(E_n-E_0)}\to0$, so $\rho\to|0\rangle\langle0|$. Since the ground state depends on external parameters, $|0\rangle=|0(\lambda)\rangle$, the infinitesimal variation takes the form
\begin{align}
\dif\rho = |\dif0\rangle\langle0| + |0\rangle\langle\dif0| + O(\me^{-\beta\Delta}),
\end{align}
where $\Delta=E_1-E_0$ is the energy gap between the ground and first excited states. The spectral variations are therefore exponentially suppressed in the zero-temperature limit, leaving only the tangential deformation associated with the ground-state manifold. Consequently, the Bures metric reduces to the Fubini-Study metric (with details explained in  Ref.~\cite{PhysRevB.110.035144}).

In contrast, the evolution according to Eq.~\eqref{Eq:rho_trans} explicitly imposes $\dif|\psi\rangle=0$ (with fixed $|\psi\rangle$ and $\epsilon\to0^+$), thereby eliminating all tangential pure-state deformations and isolating only the transverse sector. The remaining degrees of freedom are the radial parameter $\epsilon$, which controls the distance to the rank-1 pure state and the angular variables $\theta$ associated with the orthogonal complement. These are precisely the degrees of freedom responsible for the emergent conical geometry at the rank-changing point. We will explicitly show below that the Bures metric in the transverse sector indeed becomes a conical metric.

Therefore, the Fubini-Study metric describes the intrinsic geometry of the rank-1 pure-state stratum, whereas the conical metric characterizes the asymptotic transverse geometry of the surrounding mixed-state manifold near that stratum. The two structures are thus complementary rather than contradictory.
From this perspective, the analysis of the tangential sector (Fubini-Study metric) applies for any $N\ge2$, while the analysis of the transverse sector (conical metric) is intended for $N\ge3$ only. For $N=2$, transverse variations still exist but do not give rise to a genuine conical singularity because the pure-state boundary (i.e., the Bloch sphere) remains smooth. Hence the two sectors near a rank-changing point are not contradictory for $N=2$ - the transverse sector simply does not exhibit a conical structure.

\subsection{Tangential convergence to a rank-1 state}
We now provide a simplified derivation showing that the Bures metric reduces to the Fubini-Study metric under the tangential convergence to a rank-1 state. A complete proof can be found in Ref.~\cite{PhysRevB.110.035144}.

According to Eq.~(\ref{Bmetric1}), the Bures distance can be written as
\begin{align}\label{Bd1}
\dif s^2_{\mathrm B}
=
\frac12
\sum_{i,j}
\frac{
|\langle i|\dif\rho|j\rangle|^2
}{
\lambda_i+\lambda_j
},
\end{align}
where $\rho=\sum_i\lambda_i|i\rangle\langle i|$. In the zero-temperature limit, $\lambda_0=1$ and all remaining eigenvalues vanish, so $\rho\to|\psi\rangle\langle\psi|$ with $|\psi\rangle=|0\rangle$. To evaluate the Bures distance in Eq.~(\ref{Bd1}), we note
\begin{align}
\langle\psi|\dif\rho|\psi\rangle
=
\langle\psi|\dif\psi\rangle
+
\langle\dif\psi|\psi\rangle
=0.
\end{align}
For states orthogonal to $|\psi\rangle$, one finds
\begin{align}
\langle\psi|\dif\rho|a\rangle
=
\langle\dif\psi|a\rangle,
\quad
\langle a|\dif\rho|\psi\rangle
=
\langle a|\dif\psi\rangle.
\end{align}
Since $\lambda_a=0$, the denominator becomes $\lambda_0+\lambda_a=1$.

For a pure state, the terms with $i,j\ge1$ require a limiting procedure because the corresponding eigenvalues vanish. A direct evaluation shows that these terms do not contribute, while the only nonzero contributions arise from the cross terms $(0,a)$ and $(a,0)$.
Eq.~\eqref{Bd1} then gives
\begin{align}
\dif s^2_{\mathrm B}
=
\frac12
\sum_a
\left[
|\langle\psi|\dif\rho|a\rangle|^2
+
|\langle a|\dif\rho|\psi\rangle|^2
\right]
=
\sum_a
|\langle a|\dif\psi\rangle|^2.
\end{align}
Using the completeness relation,
$I=|\psi\rangle\langle\psi|+
\sum_a|a\rangle\langle a|$,
we obtain
\begin{align}
\sum_a
|\langle a|\dif\psi\rangle|^2
&=
\langle\dif\psi|\dif\psi\rangle
-
|\langle\psi|\dif\psi\rangle|^2.
\end{align}
Hence,
\begin{align}
\dif s^2_{\mathrm B}
=
\langle\dif\psi|\dif\psi\rangle
-
|\langle\psi|\dif\psi\rangle|^2
=
\dif s^2_{\mathrm{FS}}.
\end{align}
This geometry is intrinsic to the rank-1 manifold $\mathcal D_1 \cong \mathbb{CP}^{N-1}$ and describes motion tangent to the pure-state sector, including the $T\rightarrow 0$ limit of thermal states.

\subsection{Reduction of the Bures Metric to a Conical metric}\label{appc}
To prove that near a pure state the Bures metric reduces to the conical form (\ref{ds2du2}) in the transverse sector for $N\ge 3$, we parametrize the density matrices as
\begin{align}
\rho(\epsilon,\theta) = (1-\epsilon)|\psi\rangle\langle\psi| + \epsilon\sum_{a=1}^{N-1}\mu_a\,|a(\theta)\rangle\langle a(\theta)|,
\end{align}
with $0<\epsilon<1$, $\mu_a>0$, $\sum_a\mu_a=1$, and $\{|a(\theta)\rangle\}$ an orthonormal basis of the orthogonal complement of $|\psi\rangle$, varying smoothly with parameters $\theta$. Here $\mu_a$ are independent of $\epsilon$. The physical angular degrees of freedom are parametrized by the base manifold $\Sigma$, which is the space of pure states in the orthogonal complement, i.e., $\Sigma \cong \mathbb{CP}^{N-2}$. Its dimension is
\begin{align}
d = 2(N-2).
\end{align}
Thus the coordinates $\theta^a$ ($a=1,\dots,d$) denote a set of independent angular coordinates on $\Sigma$.
In this case, the  Bures distance can be expressed as
\begin{align}
\dif s^2 = \frac14\sum_{i=0}^{N-1}\frac{(\dif\lambda_i)^2}{\lambda_i}
          + \frac12\sum_{0\le i<j\le N-1}\frac{|\langle i|\dif\rho|j\rangle|^2}{\lambda_i+\lambda_j},
\end{align}
where the eigenvalues are $\lambda_0=1-\epsilon$ and $\lambda_a=\epsilon\mu_a$ ($a=1,\dots,N-1$), with eigenstates $|0\rangle=|\psi\rangle$ and $|a\rangle=|a(\theta)\rangle$.

Varying only $\epsilon$ (i.e., setting $\dif\theta=0$) gives the radial part
\begin{align}\label{radial}
\dif s^2_{\mathrm{eig}} =& \frac14\left(\frac{\dif\epsilon^2}{1-\epsilon}+\sum_a\frac{\mu_a^2\dif\epsilon^2}{\epsilon\mu_a}\right)
= \frac14\left(\frac{\dif\epsilon^2}{1-\epsilon}+\frac{\dif\epsilon^2}{\epsilon}\right)\notag\\
= &\frac{\dif\epsilon^2}{4\epsilon}+ \frac{\dif\epsilon^2}{4}+O(\epsilon).
\end{align}
Next, we evaluate the angular part. Since $|\psi\rangle$ is fixed, $\dif\psi=0$. Using $\langle a|\dif b\rangle = -\langle\dif a|b\rangle$ and $\langle\psi|\dif a\rangle=0$, we find
\begin{align}
\langle\psi|\dif\rho|a\rangle =& (1-\epsilon)\langle\dif\psi|a\rangle + \epsilon\mu_a\langle\psi|\dif a\rangle =0,\notag\\
\langle a|\dif\rho|b\rangle =&  \epsilon(\mu_b-\mu_a)\langle a|\dif b\rangle \quad (a\neq b).
\end{align}
This implies that the $(0,a)$ pairs contribute nothing and the $(a,b)$ pairs ($a,b\ge1$, $a\neq b$) lead to
\begin{align}
\frac{|\langle a|\dif\rho|b\rangle|^2}{\lambda_a+\lambda_b}
= \epsilon\,\frac{(\mu_a-\mu_b)^2}{\mu_a+\mu_b}\,|\langle a|\dif b\rangle|^2.
\end{align}
Summing over all unordered pairs $a<b$ and using the symmetry of the sum in the metric formula, we obtain
\begin{align}\label{an1}
\dif s^2_{\mathrm{ang}} = \sum_{a<b}\frac{|\langle a|\dif\rho|b\rangle|^2}{\lambda_a+\lambda_b}
= \epsilon\sum_{a<b}\frac{(\mu_a-\mu_b)^2}{\mu_a+\mu_b}\,|\langle a|\dif b\rangle|^2.
\end{align}
We identify the metric on the base space as
\begin{align}\label{h}
h_{ab}(\theta)\,\dif\theta^a\dif\theta^b \equiv \sum_{a<b}\frac{(\mu_a-\mu_b)^2}{\mu_a+\mu_b}\,|\langle a|\dif b\rangle|^2,
\end{align}
which is positive definite when the values of $\mu_a$ are not all equal (i.e., when $\sigma$ is not maximally mixed). Thus, the angular part is
\begin{align}\label{an2}
\dif s^2_{\mathrm{ang}} = \epsilon\,h_{ab}(\theta)\,\dif\theta^a\dif\theta^b + O(\epsilon^2).
\end{align}
Combining Eqs.~(\ref{radial}) and (\ref{an2}), we obtain, as $\epsilon\to0$,
\begin{align}
\dif s^2 = \frac{\dif\epsilon^2}{4\epsilon} + \epsilon\,h_{ab}(\theta)\,\dif\theta^a\dif\theta^b + O(\epsilon^2).
\end{align}
Introducing a new radial coordinate $u = \sqrt{\epsilon}$, we arrive at the conical metric shown in Eq.~\eqref{ds2du2} for the Bures metric around a pure state.

\section{Geodesic Equations on a Metric Cone}\label{appd}
To derive the geodesic equations on a metric cone, we recall that geodesics are curves that make the arc length stationary. Equivalently, one can minimize the energy functional
\begin{align}
E = \frac12\int g_{\mu\nu}\dot x^\mu\dot x^\nu\,\dif s,
\end{align}
with $\dot x^\mu = \dif x^\mu/\dif s$. For a metric cone, the associated Lagrangian is
\begin{align}
\mathcal{L}(u,\theta,\dot u,\dot\theta) = \frac12\bigl( \dot u^2 + u^2 h_{ab}(\theta)\dot\theta^a\dot\theta^b \bigr). \label{eq:Lag}
\end{align}
The Euler-Lagrange equations then give the geodesic equations.

For the radial coordinate $u$ one finds
\begin{align}
\frac{\partial\mathcal{L}}{\partial\dot u} = \dot u,\qquad
\frac{\dif}{\dif s}\frac{\partial\mathcal{L}}{\partial\dot u} = \ddot u,
\end{align}
and
\begin{align}
\frac{\partial\mathcal{L}}{\partial u} = u h_{ab}\dot\theta^a\dot\theta^b \equiv u|\dot\theta|_h^2,
\end{align}
where $|\dot\theta|_h^2 = h_{ab}\dot\theta^a\dot\theta^b$ is the squared angular speed measured with the base metric $h$. Hence
\begin{align}
\ddot u - u|\dot\theta|_h^2 = 0. \label{eq:radial}
\end{align}

For the angular coordinates $\theta^a$,
\begin{align}
\frac{\partial\mathcal{L}}{\partial\dot\theta^a} = u^2 h_{ab}\dot\theta^b,
\end{align}
and its derivative with respect to $s$ is
$\frac{\dif}{\dif s}\bigl(u^2 h_{ab}\dot\theta^b\bigr) = 2u\dot u h_{ab}\dot\theta^b + u^2\frac{\dif}{\dif s}\bigl(h_{ab}\dot\theta^b\bigr)$.
Using $\frac{\dif}{\dif s}(h_{ab}\dot\theta^b) = \partial_c h_{ab}\dot\theta^c\dot\theta^b + h_{ab}\ddot\theta^b$,
\begin{align}
\frac{\dif}{\dif s}\frac{\partial\mathcal{L}}{\partial\dot\theta^a} = 2u\dot u h_{ab}\dot\theta^b + u^2\bigl(\partial_c h_{ab}\dot\theta^c\dot\theta^b + h_{ab}\ddot\theta^b\bigr). \label{eq:theta_deriv}
\end{align}
On the other hand,
\begin{align}
\frac{\partial\mathcal{L}}{\partial\theta^a} = \frac12 u^2\partial_a h_{bc}\dot\theta^b\dot\theta^c. \label{eq:theta_partial}
\end{align}
Equating (\ref{eq:theta_deriv}) and (\ref{eq:theta_partial}) and dividing by $u^2$ ($u>0$) give
\begin{align}
h_{ab}\ddot\theta^b + 2\frac{\dot u}{u}h_{ab}\dot\theta^b + \partial_c h_{ab}\dot\theta^c\dot\theta^b - \frac12\partial_a h_{bc}\dot\theta^b\dot\theta^c = 0. \label{eq:theta_pre}
\end{align}
By renaming the dummy variables, $\partial_c h_{ab}\dot\theta^c\dot\theta^b = \frac12(\partial_c h_{ab}+\partial_b h_{ac})\dot\theta^b\dot\theta^c$. Therefore,
\begin{align}
\partial_c h_{ab}\dot\theta^c\dot\theta^b - \frac{\partial_a h_{bc}\dot\theta^b\dot\theta^c}{2} = \frac12\bigl(\partial_c h_{ab}+\partial_b h_{ac}-\partial_a h_{bc}\bigr)\dot\theta^b\dot\theta^c.
\end{align}
The right-hand side equals $h_{ad}\Gamma^d_{bc}(h)\dot\theta^b\dot\theta^c$, where $\Gamma^d_{bc}(h)$ is the Christoffel symbol of $h$. Substituting into Eq.~(\ref{eq:theta_pre}) and contracting with $h^{ae}$ yield the angular geodesic equation
\begin{align}
\ddot\theta^e + 2\frac{\dot u}{u}\dot\theta^e + \Gamma^e_{bc}(h)\dot\theta^b\dot\theta^c = 0, \label{eq:angular_full}
\end{align}
which can be rewritten as
\begin{align}
\frac{\dif}{\dif s}\bigl(u^2\dot\theta^a\bigr) + u^2\Gamma^a_{bc}(h)\dot\theta^b\dot\theta^c = 0. \label{eq:conservation}
\end{align}

We now analyze the behavior near the cone tip, $u\to0$. In this limit the term $2\frac{\dot u}{u}\dot\theta^e$ dominates over $u^2\Gamma^a_{bc}\dot\theta^b\dot\theta^c$ in Eq.~(\ref{eq:angular_full}) because the latter is of order $u^2$ while the former is of order $(\dot u/u)\dot\theta^e$. Consequently, Eq.~(\ref{eq:conservation}) simplifies to $\frac{\dif}{\dif s}(u^2\dot\theta^a)\approx 0$, so $u^2\dot\theta^a$ is approximately conserved. We define $c^a = u^2\dot\theta^a$. If $c^a\neq0$, then $\dot\theta^a = c^a/u^2$, which gives $|\dot\theta|_h^2 \sim 1/u^4$. Substituting into the radial equation (\ref{eq:radial}) yields $\ddot u \sim 1/u^3$, requiring an infinite acceleration at the tip. Since physical Lindblad evolution should guarantee finite $|\dot\psi|$ (and therefore finite angular speed), $c^a$ must be zero. Hence $\dot\theta^a=0$, and the trajectory becomes asymptotically radial.

For a purely radial evolution ($\dot\theta^a=0$) the radial equation reduces to $\ddot u=0$, implying $u(s)\sim s$. Thus,  $\lambda_{\min}=u^2 \sim s^2$.

In summary, the full geodesic equations on a metric cone are
\begin{align}
\ddot u - u|\dot\theta|_h^2 = 0,\quad
\frac{\dif}{\dif s}(u^2\dot\theta^a) + u^2\Gamma^a_{bc}(h)\dot\theta^b\dot\theta^c = 0,
\end{align}
and physical evolution that approaches the cone tip must be purely radial, $\dot\theta^a=0$. In the $u\rightarrow 0$ limit, they reduce to Eq.~\eqref{conegeo}.

\section{Curvature of the Conical Metric}\label{appe}
We consider the conical metric shown in Eq.~(\ref{ds2du2}), where the indices $a,b$ run from $1$ to $d$ and the total dimension of the cone is $d+1$. The metric components are $g_{uu}=1$, $g_{ua}=0$, and $g_{ab}=u^2 h_{ab}$. The inverse metric is $g^{uu}=1$, $g^{ua}=0$, and $g^{ab}=u^{-2}h^{ab}$, with $h^{ab}$ the inverse of $h_{ab}$.
The Christoffel symbols are given by the standard formula
\begin{align}\label{CS}
\Gamma^{\lambda}_{\mu\nu}=\frac{1}{2}g^{\lambda\rho}\left(\partial_\mu g_{\nu\rho}+\partial_\nu g_{\mu\rho}-\partial_\rho g_{\mu\nu}\right).
\end{align}
The mixed radial-angular components $\Gamma^{u}_{ab}$ are obtained by taking $\lambda=u$, $\mu=a$, $\nu=b$. Since $g_{au}=g_{bu}=0$ and $\partial_u g_{ab}=2u h_{ab}$, we find
$\Gamma^{u}_{ab}=-u h_{ab}$.
Using $g^{ac}=u^{-2}h^{ac}$, $g_{uc}=g_{ub}=0$, and $\partial_u g_{bc}=2u h_{bc}$, we obtain
\begin{align}
\Gamma^{a}_{ub}=\frac{1}{2}u^{-2}h^{ac}\cdot 2u h_{bc}=u^{-1}\delta^{a}_{b}.
\end{align}
Finally, the purely angular components $\Gamma^{a}_{bc}$ follow from $g_{cd}=u^2 h_{cd}$ and the fact that $u$ does not depend on the angular coordinates. Substituting into the definition, the factor $u^2$ cancels with $u^{-2}$ from the inverse metric, giving
\begin{align}
\Gamma^{a}_{bc}=\frac{1}{2}h^{ad}\left(\partial_b h_{cd}+\partial_c h_{bd}-\partial_d h_{bc}\right)=\Gamma^{a}_{bc}(h),
\end{align}
which are precisely the Christoffel symbols of the base metric $h_{ab}$.

The Riemann tensor is defined as
\begin{align}
R^{\rho}_{\ \sigma\mu\nu} = \partial_\mu \Gamma^{\rho}_{\nu\sigma} - \partial_\nu \Gamma^{\rho}_{\mu\sigma} + \Gamma^{\rho}_{\mu\lambda}\Gamma^{\lambda}_{\nu\sigma} - \Gamma^{\rho}_{\nu\lambda}\Gamma^{\lambda}_{\mu\sigma}.
\end{align}
Using the Christoffel symbols derived above, we compute the angular components
\begin{align}
R^{a}_{\ bcd} = R^{(h)a}_{\ \ bcd} - \delta^a_c h_{db} + \delta^a_d h_{cb},
\end{align}
where $R^{(h)a}_{\ \ bcd}$ is the Riemann tensor of the base metric $h$. Lowering the indices leads to 
\begin{align}\label{R1}
R_{abcd} =&u^2 h_{a\mu}\big( R^{(h)\mu}_{\ \ bcd} - \delta^\mu_c h_{db} + \delta^\mu_d h_{cb} \big) \notag\\=& u^2\big( R^{(h)}_{abcd} - (h_{ac}h_{bd} - h_{ad}h_{bc}) \big).
\end{align}
Next, we compute the mixed components, such as $R^u_{\ abc}$ and $R^u_{\ aub}$, and those components containing one or two $u$ indices vanish. For instance,
$R^u_{\ abc} = 0$, $R^u_{\ aub}=0$,
and lowering the indices gives
$R_{uabc}=0$, $R_{uaub}=0$.
Similarly, $R_{aubc}=0$ follows from symmetry. Thus, only the angular components of the Riemann tensor are non-zero.

The Ricci tensor is defined as $R_{\mu\nu}=R^{\rho}_{\ \mu\rho\nu}$. Its angular part is
\begin{align}
R_{ab} = R^{u}_{\ a u b} + \sum_{c} R^{c}_{\ a c b}.
\end{align}
The first term $R^{u}_{\ a u b}=0$. For the second term, using Eq.~(\ref{R1}), we have
$R^{c}_{\ a c b} = R^{(h)c}_{\ \ a c b} - \delta^c_c h_{ba} + \delta^c_b h_{ca}$.
Summing over $c$ yields
\begin{align}
\sum_c R^{c}_{\ a c b} =& \sum_c R^{(h)c}_{\ \ a c b} - \sum_c h_{ba} + \sum_c \delta^c_b h_{ca} \notag\\=& R^{(h)}_{ab} - d\,h_{ba} + h_{ba}.
\end{align}
Since $h$ is symmetric, we obtain
\begin{align}\label{R3}
R_{ab} = R^{(h)}_{ab} - (d-1)h_{ab}.
\end{align}
The radial component is
\begin{align}
R_{uu} = R^{u}_{\ u u u} + \sum_c R^{c}_{\ u c u}.
\end{align}
The first term vanishes. For the second term, we substitute the expressions from Eq.~(\ref{CS}) and use $\Gamma^{c}_{c u}=1/u$, $\Gamma^{\lambda}_{u u}=0$, $\Gamma^{c}_{u f}=(1/u)\delta^c_f$, and $\Gamma^{f}_{c u}=(1/u)\delta^f_c$ to obtain
\begin{align}
R^{c}_{\ u c u} = 0 - \partial_u\left(\frac{1}{u}\right) + 0 - \sum_f \frac{1}{u^2}\delta^c_f\delta^f_c=0.
\end{align}
Hence all $R^{c}_{\ u c u}=0$, and consequently
\begin{align}\label{R4}
R_{uu}=0.
\end{align}
Combining Eqs. (\ref{R4}) and (\ref{R3}), the scalar curvature is
\begin{align}\label{R5}
R = g^{uu}R_{uu} + g^{ab}R_{ab} =\frac{1}{u^2}\big( R_h - d(d-1) \big),
\end{align}
where $R_h = h^{ab}R^{(h)}_{ab}$ is the scalar curvature of the base manifold $h$. Since $R_h \neq d(d-1)$ in general, the scalar curvature thus diverges as $u\to0$, giving rise to a genuine intrinsic curvature singularity. 

The special case $R_h = d(d-1)$ corresponds to a flat cone with a vanishing scalar curvature $R$, but it is of very limited relevance in our discussion. To see what happens in this special case, consider $d=2$ and suppose the base manifold $\Sigma$ is a 2-sphere of radius $\kappa$, i.e.,
\begin{align}
h_{ab}\,\dif\theta^a\dif\theta^b = \kappa^2\bigl(\dif\theta^2+\sin^2\theta\dif\phi^2\bigr).
\end{align}
A straightforward computation gives its scalar curvature. Since the metric is conformal to the unit sphere, the Riemann tensor scales as $\kappa^2$, and one finds
$R^{(h)}_{\theta\phi\theta\phi} = \kappa^2\sin^2\theta$,
$R^{(h)}_{\theta\theta}=1$, and
$R^{(h)}_{\phi\phi}=\sin^2\theta$.
Hence the scalar curvature of the base manifold is
\begin{align}
R_h = h^{\theta\theta}R^{(h)}_{\theta\theta}+h^{\phi\phi}R^{(h)}_{\phi\phi}
= \frac{1}{\kappa^2} + \frac{1}{\kappa^2} = \frac{2}{\kappa^2}.
\end{align}
Therefore, the special condition $R_h = d(d-1)=2$ is satisfied when $\kappa=1$ (the unit sphere). In that case the conical metric becomes
\begin{align}
\dif s^2 = \dif u'^2 + u'^2\bigl(\dif\theta^2+\sin^2\theta\dif\phi^2\bigr),
\end{align}
which is exactly the metric of three-dimensional Euclidean space in spherical coordinates $(u',\theta,\phi)$. In this sense the ``cone'' is not a cone at all because its tip at $u'=0$ is merely the coordinate origin, and the base $S^2$ encloses the full solid angle $4\pi$.

This above special case does not arise in the Bures geometry of interest. For the $N=3$ systems with the metric shown in Eq.~(\ref{cone_metric3D}), the coefficient $\kappa^2 = \frac{(1-\zeta)^2}{8(1+\zeta)^2}$ is fixed by the spectral ratio $\zeta$ and equals unity only for the isolated value $\zeta = 1-2\sqrt{2}$. However, such a situation lies outside the physical domain $\zeta>0$. Hence for generic $\zeta$, one has $\kappa\neq1$. Consequently, $R_h = 2/\kappa^2 \neq 2$, and the scalar curvature $R\sim u'^{-2}$ diverges as $u'\to0$. The same conclusion holds for higher-dimensional base manifolds, such as $\mathbb{CP}^{N-2}$ with the Fubini-Study metric, where $R_h \neq d(d-1)$ generically.

We remark that the above derivation assumes the base manifold $\Sigma$ has dimension $d\ge2$. For $d=1$, the cone is two-dimensional, so in that case the curvature is concentrated at the tip as a Dirac delta function, which is not captured by the local curvature formula (\ref{R5}). The Gauss-Bonnet theorem provides a more appropriate description (see the two-dimensional cone example in the main text in Sec.~\ref{Sec2DCone}).

%

\end{document}